\documentclass[preprint,5p,times,twocolumn,number]{elsarticle}
\biboptions{sort&compress}
\usepackage{amssymb}
\usepackage{amsmath}
\usepackage{lipsum}
\usepackage[T1]{fontenc}
\usepackage[utf8]{inputenc}

\usepackage[colorlinks=true,allcolors=blue]{hyperref}

\journal{Physics Letters B}

\begin{document}

\begin{frontmatter}
\title{Uncertainty quantified three-body model applied to the two-neutron halo $^{22}$C }

\author[add1]{Patrick McGlynn}
\ead{mcglynn@frib.msu.edu}
\address[add1]{Facility for Rare Isotope Beams, Michigan State University, East Lansing, Michigan, USA 48824}
\author[add2,add1,add3]{Chlo\"e Hebborn}
\address[add2]{Université Paris-Saclay, CNRS/IN2P3, IJCLab, 91405 Orsay, France}
\address[add3]{Department of Physics and Astronomy, Michigan State University, East Lansing, Michigan 48824, USA}
\ead{hebborn@ijclab.in2p3.fr}

\begin{abstract}
Two-neutron  halo nuclei offer a fascinating probe into the behaviour of quantum few-body systems at the limits of binding. Although few nuclei have already been clearly identified,  many of their properties remain poorly constrained.  For example, $^{22}$C, one of the heaviest, still lacks a precise identification of its static and dynamic properties, such as its mass and dipole strength in the continuum.
One main difficulty is that  properties of two-neutron halo nuclei are inferred from indirect experimental data using a theoretical model.
Therefore, accurately determining the characteristics of two-neutron halo nuclei requires an accurate theoretical model and careful quantification of the uncertainties.
In this work, we examine $^{22}$C with a three-body model, seeing $^{22}$C  as a $^{20}$C core and two halo neutrons, and quantify for the first time the uncertainties associated with the $^{20}$C-$n$ interaction using a Bayesian approach. We propagate these uncertainties to properties of  bound and scattering states of $^{22}$C, as well as its dipole strength.   The comparison of our prediction for the matter radius to experimentally-derived  values suggests that $^{22}$C is bound by less than 0.35~MeV and is dominated by a $(s_{1/2})^2$ configuration.  Our analysis of the dipole strength shows (i) that final-state interaction needs to be included for an accurate description, (ii) the uncertainties on the dipole strength function are about 50\% and are mostly influenced by uncertainties on the  $^{22}$C ground-state  properties, i.e. its binding energy and single-particle structure, and (iii)  partial-wave occupation of $^{22}$C depends on  the scattering length and the $d_{3/2}$ resonance energy of the $^{20}$C-$n$ unbound system. Such enhanced sensitivity of the dipole strength to  both $^{21}$C  and $^{22}$C properties motivates  a precise measurement of the  $^{22}$C  dipole strength function, that will allow to precisely and accurately resolve the spectroscopy of these nuclei.

\end{abstract}


\end{frontmatter}
\section{Introduction and motivation}
Recent experimental progress has allowed for measurements of isotopes near the driplines, exhibiting exotic features such as unexpected shell ordering \cite{Fortunato2020,Singh2024a,Poves2017} and clusterized structures~\cite{Bazin2023,Freer2018}, thereby challenging traditional models of nuclear structure. Among the most extreme manifestation of clusterizing  are halo nuclei~\cite{Tanihata2013}, in which one or more loosely bound nucleons are found in spatially diffuse wavefunctions extending far beyond the core. Two-neutron halo nuclei, i.e. those composed of a core and two halo neutrons, can exhibit a Borromean structure, meaning that the three-body system forms a bound state, but none of the two-body subsystems is bound. These Borromean structures are strongly influenced by  couplings to the continuum and offer an interesting probe of strongly correlated few-body systems including the dineutron \cite{Casal2019a,Lovell2017,Costa2025,Monteagudo2024}. The identification of such exotic structures relies on state-of-the-art experimental techniques and robust theoretical models to interpret the data. One particularly interesting observable is the dipole strength distribution in the continuum, which  exhibits a characteristic low-energy enhancement~\cite{Aumann2013}.  These dipole strengths are commonly inferred from Coulomb breakup measurements and have been studied  for   light halo nuclei, including $^6$He~\cite{Aumann1999a,Wang2002a}, $^{11}$Li~\cite{Nakamura2006a}, $^{14}$Be~\cite{Labiche2001}, $^{19}$B~\cite{Cook2020} $^{22}$C~\cite{Kobayashi2012,Nagahisa2018} and $^{29}$F~\cite{Bagchi2020}. From this dipole strength, one can extract, provided an accurate theoretical model, key properties of the halo systems, such as its binding energy. However, this interpretation is inherently model dependent and therefore introduces systematic uncertainties that must be carefully quantified.

Recent progress in \textit{ab initio} nuclear theory has allowed structure calculations to reach the dripline, describing one-nucleon and even some two-neutron halo nuclei~\cite{navratil_ab_2011,Calci2016,reviewncsmc,quaglioni_three-cluster_2018,Letter6HeNCSMC,Kravvaris8B,navratil_halo_2026,elhatisari_ab_2017,song_ab_2026,shen_ab_2025,Bonaiti2022,Hagen2010b,Hagen2013,Shen2026,Li2024b,Rodkin2021,cxcz-z8px,PhysRevLett.111.132501,SANCSMHalo,Huang2026,Hagen2013a,Hiyama2022,Hiyama2019}. However, these approaches remain computationally demanding and have therefore only been applied to a limited number of systems. As a result, properties of two-neutron halo nuclei  have typically been predicted using three-body models \cite{Nunes1996a,Tostevin2001,Thompson2004c,Lovell2017,Pinilla2016,Pinilla2025,Descouvemont2003,Descouvemont2006,Casal2019a,Casal2020a,Ershov2012,Horiuchi2006} which use the core and valence neutron degrees of freedom and   accurately include couplings to continuum states.  
These models rely on effective core-neutron interaction, which are usually poorly-constrained due to the scarcity of experimental data. Provided that the three-body problem is solved accurately most of the uncertainties stem from these interactions.
Because halo systems exhibit a natural separation of  scales between the core and valence nucleons, they are ideal candidate for an effective field theory (EFT) description \cite{Bertulani2002,Hammer2017a,Hammer2020,Capel2021,Braun17C}. Such a model uses only a few low-energy constants, often the scattering lengths and binding energies. At leading order, only the binding energy is used to constrain the EFT and the properties of the system are described by a universal behavior~\cite{Gobel2024,Hongo2022,Acharya2013}. One main advantage of such approach is that the  uncertainties associated with the truncation of the EFT can be quantified.  
Nevertheless, these models typically focus on a few low-$l$ orbitals, which makes it challenging to capture the structure of higher-$l$ shells in the nucleus.  
With more experimental data becoming available in the mid-mass region, halo structures are expected to be found around deformed cores, for which an accurate description of higher-$l$ orbitals will be needed.

$^{22}$C is one of the heaviest confirmed two-neutron halos, and besides its existence, little is known for sure about it. Direct mass measurements \cite{Gaudefroy2012} and mass evaluations \cite{Wang2021} only set an upper bound for the two-neutron separation energy of $\sim$500~keV. Two independent measurements of the total interaction cross section for $^{22}$C suggested a large matter radius, but led to discrepant values \cite{Tanaka2010,Togano2016}. In terms of its shell structure, in the absence of low-energy continuum states in the $^{21}$C system, one expects that core neutrons fill the $d_{5/2}$ level and the two halo neutron would be in a $s$-wave.  
However, experimental studies suggest the presence of 
a virtual $s$-wave state, which could enhance halo formation \cite{Mosby2013}, as well as a low-energy $d_{5/2}$ resonance~\cite{Leblond2015}. 
To further resolve the properties of $^{21}$C and $^{22}$C, it is essential to incorporate all available experimental information within a robust statistical framework, thereby improving the theory–experiment comparison used to refine their spectroscopy.

In this work, we quantify for the first time the uncertainties associated with the calibration of the  effective $^{20}$C-$n$ interaction using a Bayesian approach, and propagate those uncertainty to $^{22}$C structure and reaction observables. 
We work within a framework of hyperspherical harmonics with an R-matrix approach \cite{McGlynn2026}, which  treats consistently bound and scattering states. Using our uncertainty-quantified prediction, we investigate the competition between universal behaviour and single-particle structure of $^{22}$C . We detail the Bayesian calibration of the $^{20}$C-$n$ interaction in  Sec.~\ref{UQ}. We present in Sec.~\ref{22Cobservables} our predictions for $^{22}$C observables, and compare to the matter radius derived from experimental data~\cite{Tanaka2010,Togano2016}. We also  discuss how a precise measurement of  $^{22}$C dipole strength   could be used to determine precisely the two-neutron separation energy and the single-particle structure of $^{22}$C as well as the low-lying continuum states of $^{21}$C. Sec.~\ref{conclusions} contains the conclusions and prospects of this work.

\section{Calibration of two- and three-body parameters}
\label{UQ}
In our three-body model, we describe $^{22}$C as composed of an inert $^{20}$C core in its $J^\pi=0^+$ ground state, to which two neutrons are loosely bound. We use a Hamiltonian composed of two-body interactions ($^{20}$C-$n$ and $n$-$n$) and three-body interactions (more details about the formalism can be found in Ref.~\cite{McGlynn2026}). The $n$-$n$ interaction used is the Minnesota potential \cite{Thompson1977,Varga1995,Bogner2011}, which reproduces low-energy $n$-$n$ scattering. The $^{20}$C-$n$ interaction is  parametrised  as  a Woods-Saxon potential defined for each partial wave by a depth $V_l$ and global radius $R$ and diffuseness $a$. We calibrate  the core-neutron potential parameters $V_s$ and $V_d$ (the depths of the $s$- and $d$-wave potentials), $V_{ls}$ (the strength of the spin-orbit force) and $R$ (the Woods-Saxon range), to the limited available information  on  $^{21}$C low-lying spectrum using a Bayesian approach.

 In the $s$-wave, we consider the existence of a virtual state\footnote{Note that the authors of \cite{Leblond2015} suggest an $s$-wave resonance to account for the observed $s$-wave peak at low energy. However, in our three-body model the introduction of such a resonance leads to a highly compact system incompatible with a halo structure.}, as suggested in Ref.~\cite{Mosby2013}.  In particular we use for the scattering length of this virtual state $-2.8 \pm 1.4 $~fm, using as mean value  the one determined in \cite{Mosby2013}. Since the original work did not provide any uncertainty, we take a conservative approach and assign a large  error to this scattering length. For the $d$-wave, we consider the $d_{5/2}$ state to be bound by $1.5\pm 0.1$~MeV, consistent with the peak of the $n$-$^{20}$C relative energy distribution seen in Ref. \cite{Leblond2015}, and following their interpretation of that $l=2$ peak.We treat the $d_{5/2}$ state as completely occupied by neutrons in the core, and the $d_{3/2}$ to have no occupation by core neutrons. Since they do not identify another $l=2$ peak, the $d_{3/2}$ resonance is assumed not to be seen below $2.5$~MeV, which we also use to constrain our calibration. Finally, we require that there be exactly one bound state in the $s_{1/2}$, $p_{3/2}$ and $p_{1/2}$ waves and no $p$-wave resonance below $5$~MeV to match the structure of $^{20}$C. 
 
 We use wide Gaussian priors for the potential parameters: both $V_d$ and $V_s$ use a prior centred at $40$~MeV with a width of $20$~MeV, consistent with the range of similar model calculations~\cite{Pinilla2016}. 
 The geometry of the potential was informed by $R=1.2A^{1/3}$~fm, giving a central value of $3.26$~fm and a width of $0.75$~fm. We fix the diffuseness to $0.65$~fm so that only the range controls the long-range behaviour. Since no relevant two-body data exists to constrain the $p$-wave, $V_p$ was taken as the average of $V_d$ and $V_s$.  Finally the spin-orbit strength lacks an obvious choice of prior, so the value used in Ref. \cite{Pinilla2016} was used to give the central value of $17.2$~MeV\footnote{Our spin-orbit force is defined differently leading to a scaling by a factor of $\sim$2 compared to Ref. \cite{Pinilla2016}.} and a width of $5$~MeV was chosen. More details about the Bayesian calibration are given in \ref{sec:bayesian}.

We propagate the uncertainties of the $^{20}$C-$n$ interaction to the $^{22}$C observables by performing three-body calculations for 315 samples of our posterior distributions. As explained in Ref.~\cite{McGlynn2026}, the Pauli-forbidden states of the core-neutron potential are removed using a projection operator and we use a three-body force to fix the  two-neutron separation energy $S_{2n}$.  This quantity is not well known for $^{22}$C, with the two values $0.035\pm 0.020$~MeV (as evaluated in AME2020 \cite{Wang2021}) or $-0.14\pm 0.46$~MeV (as obtained from a direct mass measurement \cite{Gaudefroy2012}), both being consistent with a range between $0$ and $\sim 0.5$~MeV. Previous calculations~\cite{Acharya2013,Yamashita2011,Horiuchi2006} agree with this range of separation energies, and  tend to favour a smaller separation energy. To reflect this uncertainty and determine its effect on observables, we use five values of $S_{2n}$: $0.1$, $0.2$, $0.35$, $0.5$ and $1.0$~MeV. These are controlled by modifying the strength of the three-body force. Since the three-body force only applies in the $J^\pi=0^+$ channel, the $1^-$ scattering state is unchanged. In the following section, we analyse the credible intervals for $^{22}$C observables, that are computed with the model space described in Ref.~\cite{McGlynn2026}.

\section{Prediction for  $^{22}$C properties}
\label{22Cobservables}

\begin{figure}
 	\centering
 	\includegraphics[width=\linewidth]{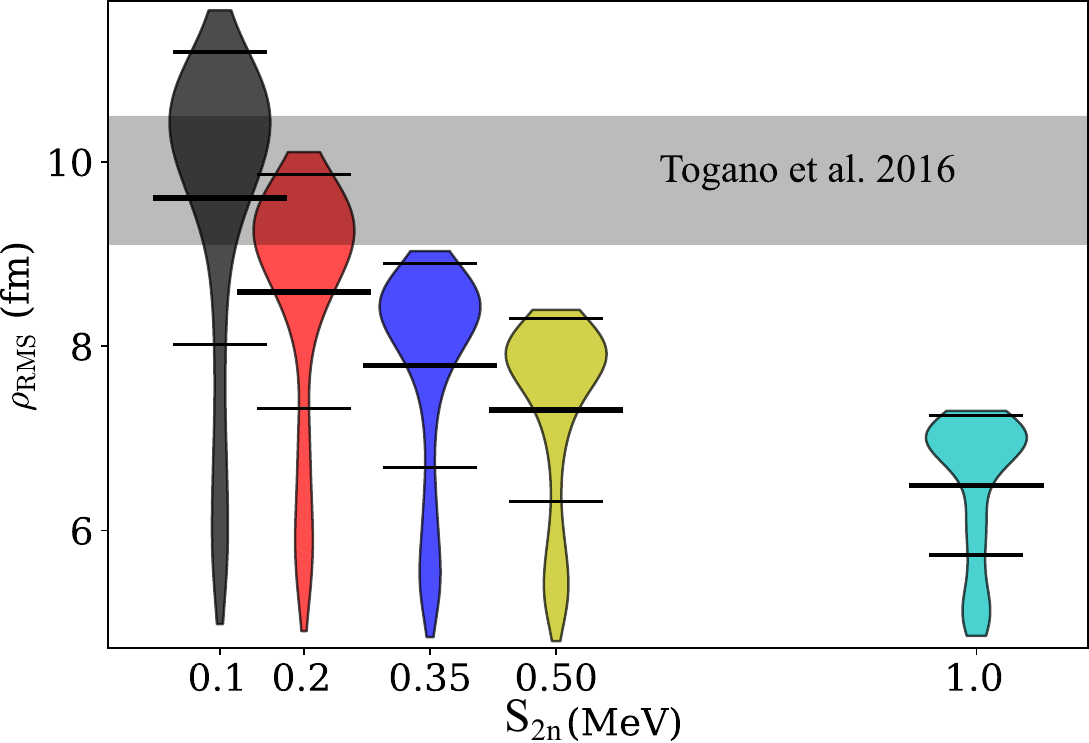}
 	\caption{Root-mean-square hyperradius as a function of two-neutron separation energy. Violin plots show a smoothed probability density function computed from 315 samples for each energy. Horizontal lines indicate the mean and $\pm$1$\sigma$ intervals. The 1$\sigma$ interval from Ref. \cite{Togano2016} is shown. Note the cutoff of the probability density function at the extreme values does not imply it is impossible to exceed these values.}
 	\label{fig:RMS}
 \end{figure}

To study how uncertainties in the $^{20}$C-$n$ interaction influence the  halo character of $^{22}$C, we naturally start by analyzing $^{22}$C root-mean-squared (rms) matter radius.
In a  three-body model of $^{22}$C, the matter radius can be separated into two components~\cite{Pinilla2016,Ershov2012}
\begin{equation}
    \langle r^2\rangle_{^{\rm 22}C}=\frac{20}{22}\langle r^2\rangle_{^{\rm 20}C}+\frac{1}{22}\langle \rho^2\rangle
\end{equation}
 where $\sqrt{\langle r^2\rangle_{^{\rm 20}C}}$ is the rms matter radius of the $^{20}$C core and $\sqrt{\langle\rho^2\rangle}\equiv \rho_{RMS}$ is the rms hyperradius of the three-body $^{20}$C+$n$+$n$ system.   
 We show in  Fig.~\ref{fig:RMS}  our predictions for the rms hyperradius  for various separation energies, with each violin shape obtained from the 315 samples of the $^{20}$C-$n$ interaction.  As expected, the radius decreases when the separation energy increases. Interestingly, for all values of $S_{2n}$, two ``modes" appear in the distribution of the rms radii: the majority of samples favour a larger hyperradius, while the minority show a preference for a much lower hyperradius, with less dependence on separation energy. Further investigation (detailed in   \ref{sec:separation}) shows that that the larger hyperradii corresponds to $s$-wave dominated samples, i.e. parameters in which the $^{20}$C-$n$ scattering length is large or the $d_{3/2}$ resonance energy is high, and the tail to smaller hyperradii correspond to the to $d$-wave dominated samples. 
 The presence of these two modes demonstrates that rms radii are  sensitive  to the single-particle  structure of halo systems.

We now compare our predicted rms hyperradii with $^{22}$C  matter radii derived from measurements of interaction cross sections on a hydrogen target
$\sqrt{\langle r^2\rangle_{^{\rm 22}C}}=5.4\pm0.9$~fm~\cite{Tanaka2010} and on a carbon target on $\sqrt{\langle r^2\rangle_{^{\rm 22}C}}=3.4\pm0.08$~fm~\cite{Togano2016}.
Because the reported uncertainties do not include contributions from theoretical modelling in the analysis of the reaction data, they are likely underestimated, which could explain the observed discrepancies~\cite{Smith2026}. 
Using the experimentally-derived value of $^{20}$C rms radius ($\sqrt{\langle r^2\rangle_{^{\rm 20}C}}=2.98\pm0.05$~fm~\cite{Ozawa2001}), these two values correspond respectively to $\rho_{RMS}=22\pm5$~fm and $\rho_{RMS}=9.1\pm 0.7$~fm.   None of our calculations in Fig.~\ref{fig:RMS} are compatible with the value derived in Ref.~\cite{Tanaka2010} unless the $^{20}$C core is assumed to have a much larger size, and only calculations reproducing $S_{2n}\lesssim 0.35$~MeV that are $s$-wave dominated
are consistent with the value of Togano \textit{et al.}~\cite{Togano2016}. This is consistent with previous predictions  rms radius of $^{22}$C~\cite{Horiuchi2006,Ershov2012}.  This suggests that the $^{22}$C ground state has a $s$-wave dominated single-particle structure and is bound by less than 350~keV.  Nevertheless, it would be worth revisiting this analysis in the future including the  uncertainties associated with the  description of the reaction with the target.

\begin{figure}
    \centering
    \includegraphics[width=\linewidth]{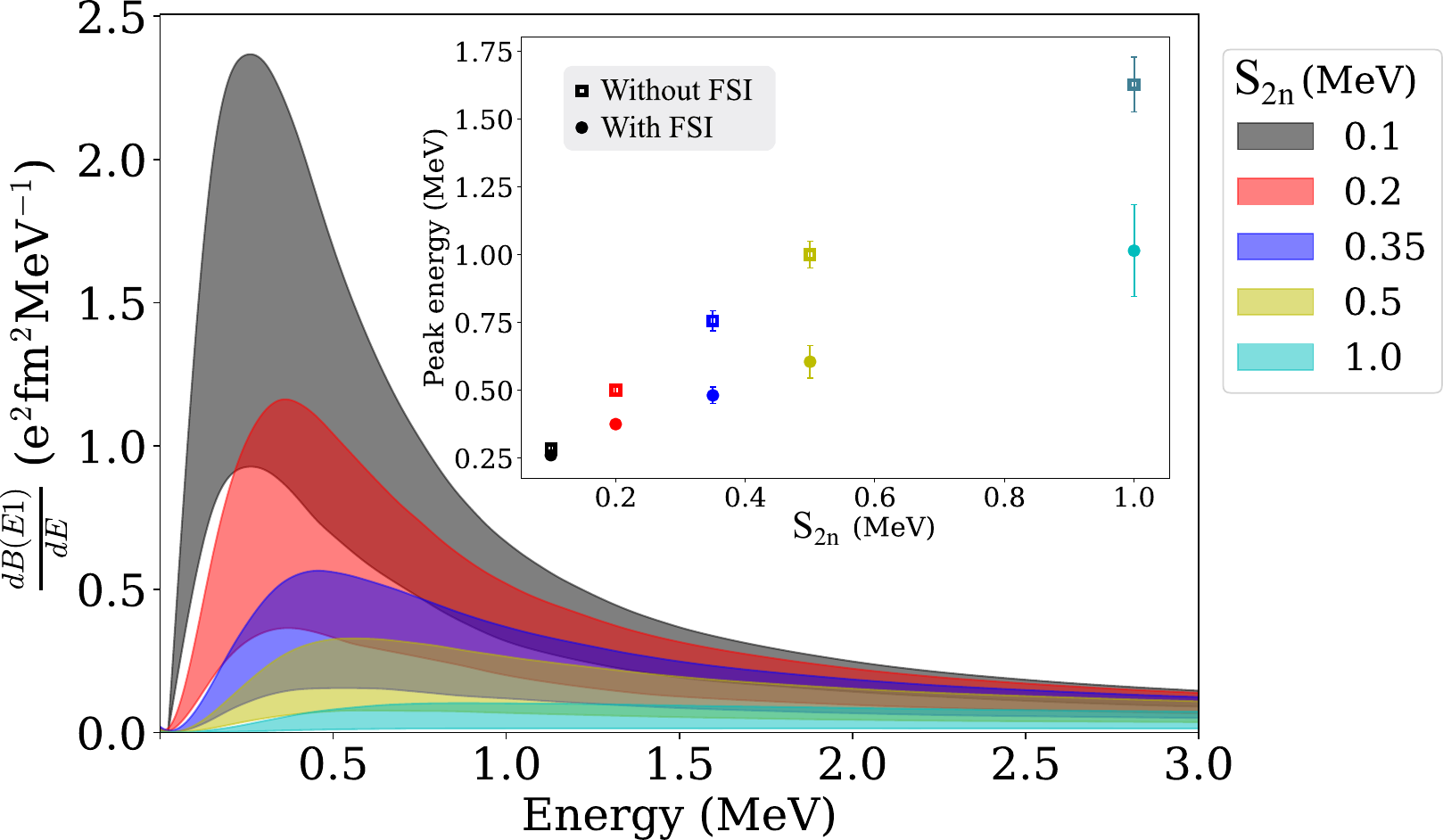}
 
    \caption{68\% credible intervals for the dipole strength function of $^{22}$C as a function of the relative energy of the three-body system. The different colours correspond to different two neutron separation energies. (Inset) Energy of the peak of the $dB(E1)/dE$ curve as a function of $S_{2n}$ with error bars indicating 68\% credible intervals. Points are computed using both unapproximated scattering waves and a plane wave approximation in the $J^\pi=1^-$ scattering channel to determine the effect of final state interactions (FSI).}
    \label{fig:UQ}
\end{figure}

As previously mentioned, another interesting observable for halo nuclei is the dipole response in the continuum. This observable describes the $E1$ excitation from the  ground state to a state in the continuum, it is hence computed from both bound and scattering  wavefunctions.  Using our three-body framework of hyperspherical harmonics with an R-matrix approach, both wavefunctions are computed consistently and with the correct boundary condition. Fig.~\ref{fig:UQ} shows our prediction for  the $dB(E1)/dE$ strength function obtained with the 315 samples of the $^{20}$C-$n$ interaction and for different $S_{2n}$ values (colours). As  expected for a halo system~\cite{Hongo2022,Acharya2013}, we see a peak at low energy, with the energy of that peak increasing as the system becomes more bound. 
As shown in the inset Fig.~\ref{fig:UQ}, the peak energy depends approximately linearly
on the three-body binding energy, as expected from a universal picture\footnote{Inspection of eqns. 29\&30 of Ref. \cite{Hongo2022} results in an almost linear dependence of the peak energy on the binding energy.}. Nevertheless, the  large errors on the slope of the line indicates deviation from this universal behavior.  In terms of magnitude, 
the overall dipole strength decreases as the separation energy increases. This can be understood considering centers of mass and charge are closer together for more bound systems, resulting in a smaller dipole strength. In terms of uncertainties, the overall scale of the credible interval  is large for all separation energies and corresponds approximately to a constant 50\% error. This indicates a dependence on the details of the $^{20}$C-$n$ potential, again suggesting deviations from a truly universal picture.

To understand the influence of the scattering states on the dipole strength, we repeated these calculations considering no final-state interactions (FSI), i.e., approximating the scattering solutions with plane waves, rather than using the full solutions to the three-body problem.  The plane-wave predictions  lead to dipole strength distribution with a peak at a significantly higher energy than the full calculations  (compare the empty squares and the filled circles in the inset of Fig.~\ref{fig:UQ}). This shift in the energy of the peak is  more pronounced  for more deeply-bound states. That  can also be intuitively understood considering the spatial extension of the two-neutron halo nucleus: a more bound system will be less spatially extended, and hence each cluster (core or halo neutron) is more sensitive to the interactions with the other clusters. Moreover, we show  in~\ref{sec:sensitivity} that although including FSI  affects the overall shape of the $dB(E1)/dE$, it does not change the uncertainties on its magnitude. This indicates that the large uncertainties in $dB(E1)/dE$  stem  from the bound-state description. 
This large sensitivity of the $dB(E1)/dE$ to the details of the bound state wavefunction confirms that this observable is a powerful tool to study the ground states of two-neutron halo systems. Nevertheless, using a model without FSI to extract properties of a two-neutron halo, such as its binding energy, from a measurement of $dB(E1)/dE$ would lead to a skewed result.

\begin{figure}
    \centering
    \includegraphics[width=\linewidth]{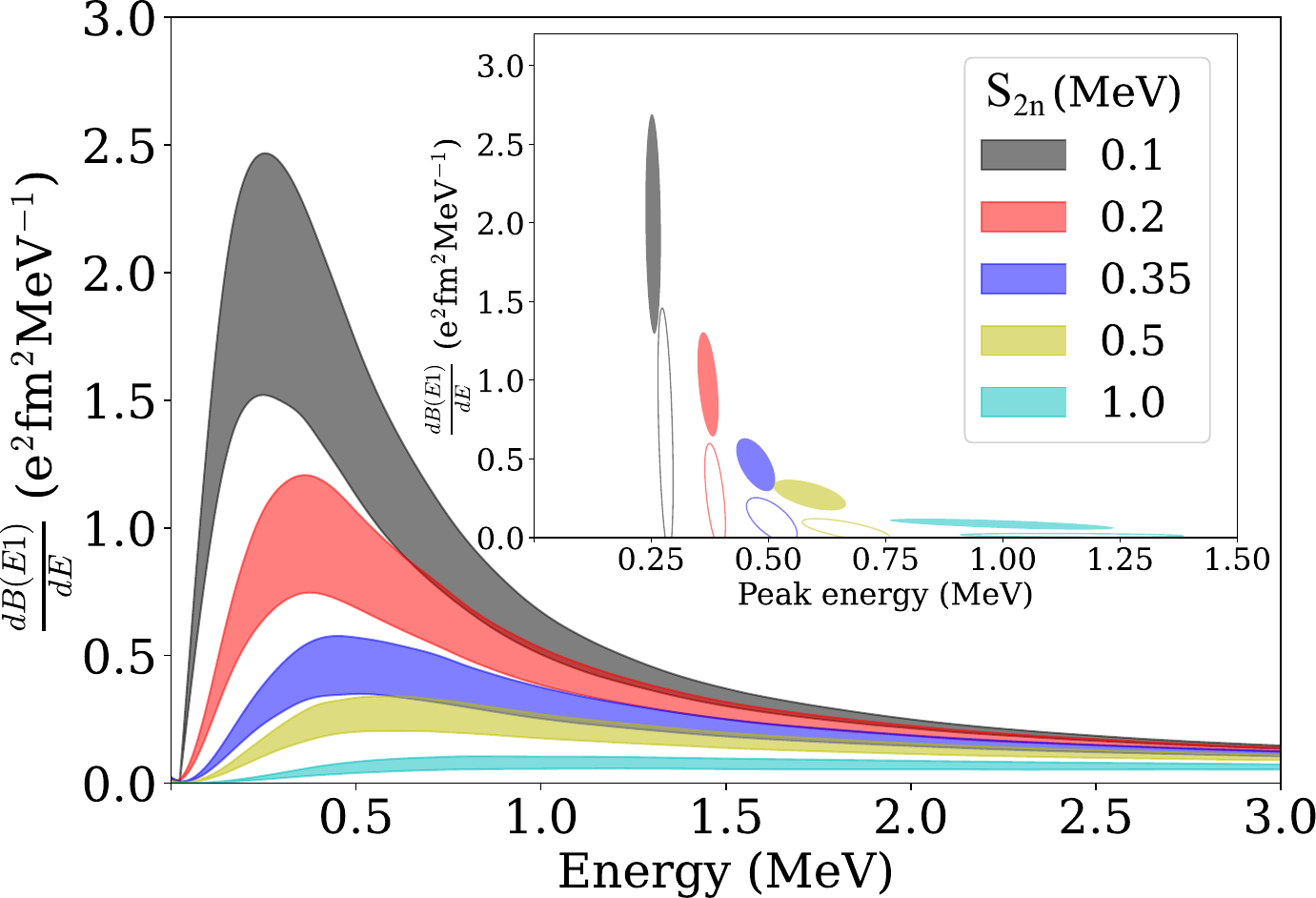}
   
    \caption{Same as Fig.\ref{fig:UQ} using only $s$-wave dominated samples. (Inset) 68\% credible ellipses for the peak energy and the value of $dB(E1)/dE$ at the peak, separated into $s$- (colored ellipse) and $d$-wave (unfilled ellipse) dominated samples.}
    \label{fig:separate}
\end{figure}

Similarly to the study on the rms hyperradius, we now investigate the sensitivity of $dB(E1)/dE$ to the single-particle structure of the ground state. Fig.~\ref{fig:separate} shows  predictions for the $dB(E1)/dE$ that corresponds to the $s$-wave dominated $^{22}$C states (colored bands).  Relative to Fig.~\ref{fig:UQ}, obtained from all $s$- and 
$d$-wave samples, credible intervals for $dB(E1)/dE$  are now much narrower and the upper limits of   $dB(E1)/dE$ are similar across  all energies. 
This reduction of uncertainties occurs because the relative error from the bound states in the $s$-wave samples is much smaller than in the $d$-wave samples (see \ref{sec:sensitivity}). The smaller amplitude of the $d$-wave dominated calculation  is explained partly by the more compact shape (seen in Fig.~\ref{fig:RMS} and caused by a larger centrifugal barrier), and partly by the fact that the $E1$ transition between the $1^-$ scattering state and $0^+$ ground state proceeds mostly to the $s$-wave component of the bound state. 
The $s$-wave curves are  more consistent with the universal picture \cite{Hongo2022,Costa2025} of a weakly-bound halo nucleus, i.e., the credible intervals in Fig.~\ref{fig:separate} can be well described by simple functions of  the binding energy.%\footnote{Specifically, the peak energy is excellently reproduced by the function $E_{peak}=0.18\text{ MeV}+0.82S_{2n}$ and the height of the peak by $dB(E1)/dE_{peak}=0.17e^2\text{fm}^2/S_{2n}$, so a scaling of the entire curve to match those values transforms between different binding energies.} 

The dipole strength shape and magnitude are strongly sensitive to both the two-neutron separation energy and the single-particle structure of $^{22}$C; a precise measurement of this observable would enable to more accurately determine $^{22}$C properties.  In the inset of Fig.~\ref{fig:separate}, we show how a comparison with an experiment could provide two simultaneous pieces of information: the separation energy of $^{22}$C from the position of the peak, and  the relative contribution of $s$- and $d$-waves in the $^{22}$C ground state from the peak height. Moreover, using our Bayesian framework, this theory-experiment comparison would also help clarify the properties of the low-lying spectrum of $^{21}$C.

\section{Conclusions}
\label{conclusions}

In this work, we have  presented the first uncertainty-quantified three-body prediction of a two-neutron halo nucleus, using   a Bayesian calibration of the core-neutron interaction. Focusing on $^{22}$C, the properties of which remain poorly constrained, we employ a Bayesian approach to calibrate $^{20}$C-$n$  interaction on  available information on $^{21}$C system and propagate these uncertainties to properties of $^{22}$C. This framework allows  us to accurately describe both the universal features and the signatures of  single-particle properties in the  two-neutron halo observables. 
 
 Our study focuses on two key observables for halo systems:  the rms matter radius and the dipole strength in the continuum.
We find that both observables are strongly sensitive to the two-neutron separation energy and the partial-wave content of the ground state wavefunction. 
Comparing our model's predictions to a matter radius derived from  interaction cross-section measurements \cite{Togano2016}, our results suggest that $^{22}$C has a two-neutron separation energy  below 0.35~MeV and a ground state dominated by a $(s_{1/2})^2$ configuration for the halo neutrons. In future work, we plan to revisit this analysis including the uncertainties associated with the analysis of the interaction cross sections, that were not included here.

 We then turn to the dipole strength, and we highlight  the importance of  including  final-state interactions for an accurate interpretation of experimental data. 
Our study also shows that dipole strengths corresponding to $s$-wave dominated  ground states exhibit the expected universal behaviour, i.e. the peak position depends linearly on the binding energy of the three-body system. This universality is lost for $d$-wave dominated ground states. 
Moreover, our Bayesian approach would  enable  back-propagation of these   $^{22}$C data to the properties of $^{21}$C, which remain poorly known, including the scattering length and the energy of the $d_{3/2}$ resonance.  Such a study would clarify the shell structure of the $sd$ levels in this region of the nuclear chart and would provide useful insights not only into this system, but also into the behaviour of loosely-bound neutron-rich nuclei in general.

Finally, this work demonstrates the viability of performing Bayesian uncertainty quantification of three-body calculations without  approximating the three-body continuum. Applying this framework to other  two-neutron halo systems would enhance theory-experiment comparisons  and  refine  the properties  of these systems inferred from data.  Future works include studying other two-neutron halo nuclei and embedding our predictions in a reaction framework to enable direct  comparison with experimental data, similarly to Ref. \cite{Pinilla2016}, rather than with derived values. This will allow  more accurate comparisons with  experiments on two-neutron halo systems.

\section*{Acknowledgements} 
We are grateful to Miguel Marques, Nigel Orr, Julien Gibelin and Daniel Phillips for insightful discussions related to this work.  
We also thank  the few-body group at MSU for regular discussions and support. This project also received financial support from the CNRS through the AIQI-IN2P3 project. Calculations were performed using the High-Performance Computing Center at MSU's Institute for Cyber-Enabled Research.

\appendix
\section{Calculation of Bayesian posterior}
\label{sec:bayesian}
\begin{figure}
    \centering
    \includegraphics[width=\linewidth]{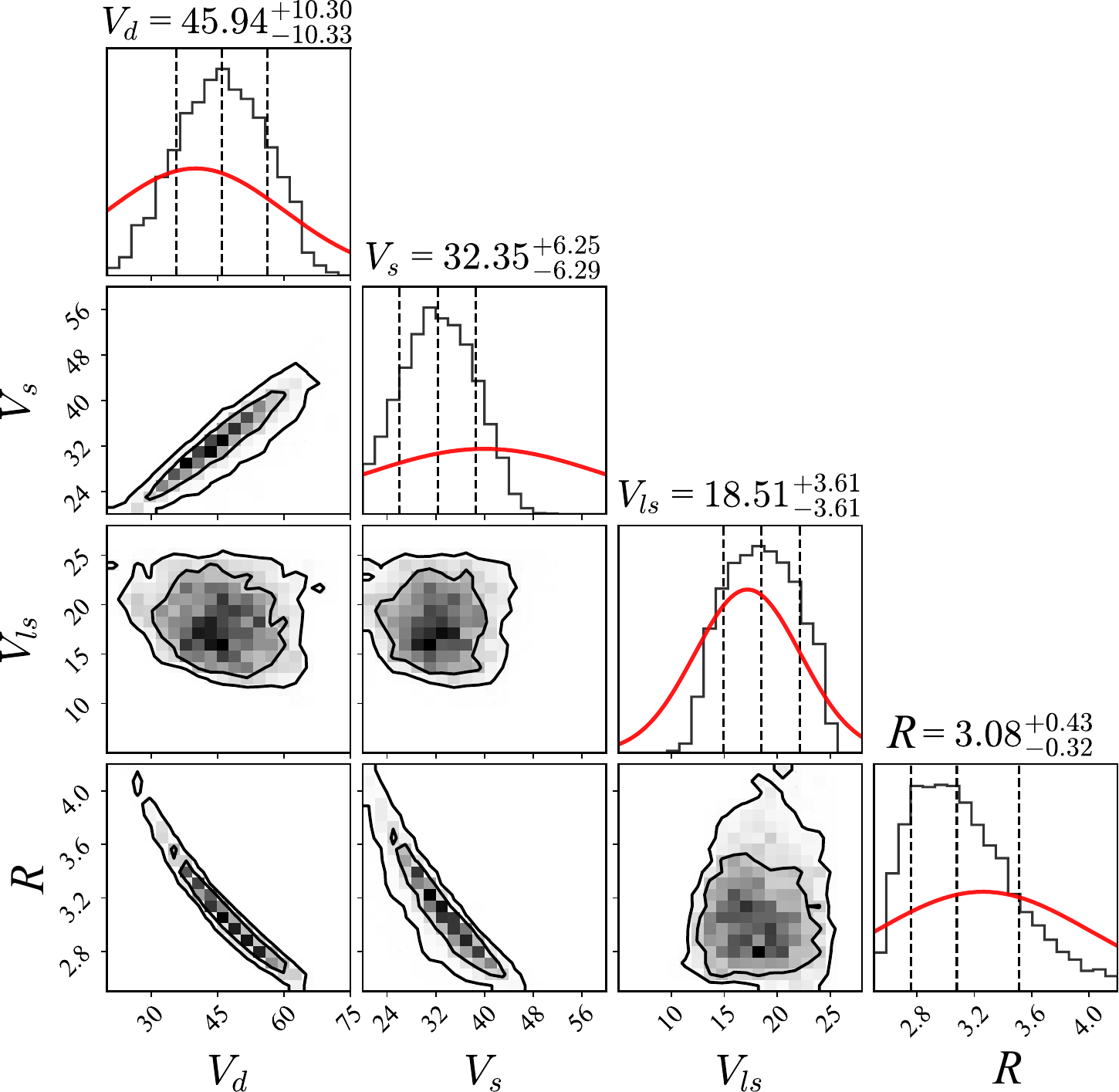}
    \caption{Two-parameter corner plots of posterior distribution of  $^{20}$C-$n$ potential parameters. Red lines show the prior distribution used in the Bayesian calibration.}
    \label{fig:cornerplot}
\end{figure}

In this appendix, we provide more details about the Bayesian calibration of the  $^{20}$C-$n$ potential parameters  and show  the corner plot of the posterior distributions (Fig.~\ref{fig:cornerplot}).   The posterior distributions are obtained via a Monte-Carlo sampling process using the python package \texttt{emcee}~\cite{emcee} and are the result of 20000 steps using 32 walkers, after discarding the first 1000 steps as burn-in. The corner plot is obtained from 10000 samples  selected  randomly from the posterior distribution to build the cornerplot. The 315 samples used for the three-body model are drawn from this subset of 10000. 

Fig.~\ref{fig:cornerplot}  shows the posterior (histograms) alongside the prior (solid red line) distributions. Except for $V_{ls}$, none of the  posterior distributions are  prior-dominated, indicating that the calibration data efficiently informs the parameters. The  influence of the  prior for $V_{ls}$  is unsurprising, given that the spin-orbit splitting can only be suitably constrained with information about both levels of a spin-orbit pair, or indirectly with enough information about resonance energies and widths in some cases. Since such information is not available, the spin-orbit parameter remains  poorly constrained and would benefit from  more precise information on  the $^{20}$C-$n$ system.

Fig.~\ref{fig:cornerplot} also shows that  several of the posterior distributions are non-Gaussian, especially the distribution of $R$, confirming   the suitability of using a Bayesian approach over simpler frequentist methods~\cite{BANDmanifesto,Pruitt2024}. The apparent correlation between $s$ and $d$ wave depths $V_s$ and $V_d$ arises from the imposition of the same geometry in both partial waves. 
In all future calculations, unless stated otherwise, a random selection of 315 such samples is used, with that number being driven by the need for convergence of the \textit{three-}body credible intervals.

\section{Separation of calculations into $s$-wave and $d$-wave dominated samples}

\label{sec:separation}

As shown in Fig.~\ref{fig:RMS}, part of the three-body calculations predicts  larger rms hyperradii, while another part yields smaller hyperradii; this behavior is observed across all separation energies. To better understand the origin of these two groups, we  focus on the calculations reproducing $S_{2n}=0.5$~MeV. We divide  the samples into two categories: the ones that lead to rms hyperradii above 6.5~fm (246 out of 315 samples considered in this work) and the remaining samples, producing smaller rms hyperradii.  The threshold of 6.5~fm corresponds approximately to the minimum of the probability distribution between the two modes. Interestingly, calculations from each group lead to $^{22}$C ground states that have similar partial-wave decompositions.

 Before presenting in details the partial-wave decomposition obtained for  each group, we  clarify the model employed in this  work. Within our three-body model of two-neutron halo nucleus,  two  Jacobi sets are relevant  \cite{McGlynn2026}. The first, which we refer to as the Y-basis\footnote{In practice there are two Y-bases, one for each neutron, but the indistinguishability of neutrons allows us to describe only one. } (shell-model like), is defined by the core-neutron relative coordinate  $\vec a$ and  the coordinate from the other neutron to the center of mass of the core-neutron subsystem $\vec b$. The second set, the T-basis, is defined by neutron-neutron and the core-dineutron relative coordinates $\vec x$  and $\vec y$, respectively.
Each basis carries its own set of orbital angular momenta: $[l_a,l_b]$   in the Y-basis and $[l_x,l_y]$ in the T-basis. The same state  with the total angular momentum and parity $J^\pi$ has a different decomposition in terms of partial waves in each basis. 
In the specific case of $^{22}$C, modeled as a $^{20}$C core  in its $0^+$ ground state with two halo neutrons, we predict  a $J^\pi=0^+$ ground state and  partial waves with $l_x=l_y$ and $l_a=l_b$ are populated.

\begin{figure}[t]
	\centering
	\includegraphics[width=\linewidth]{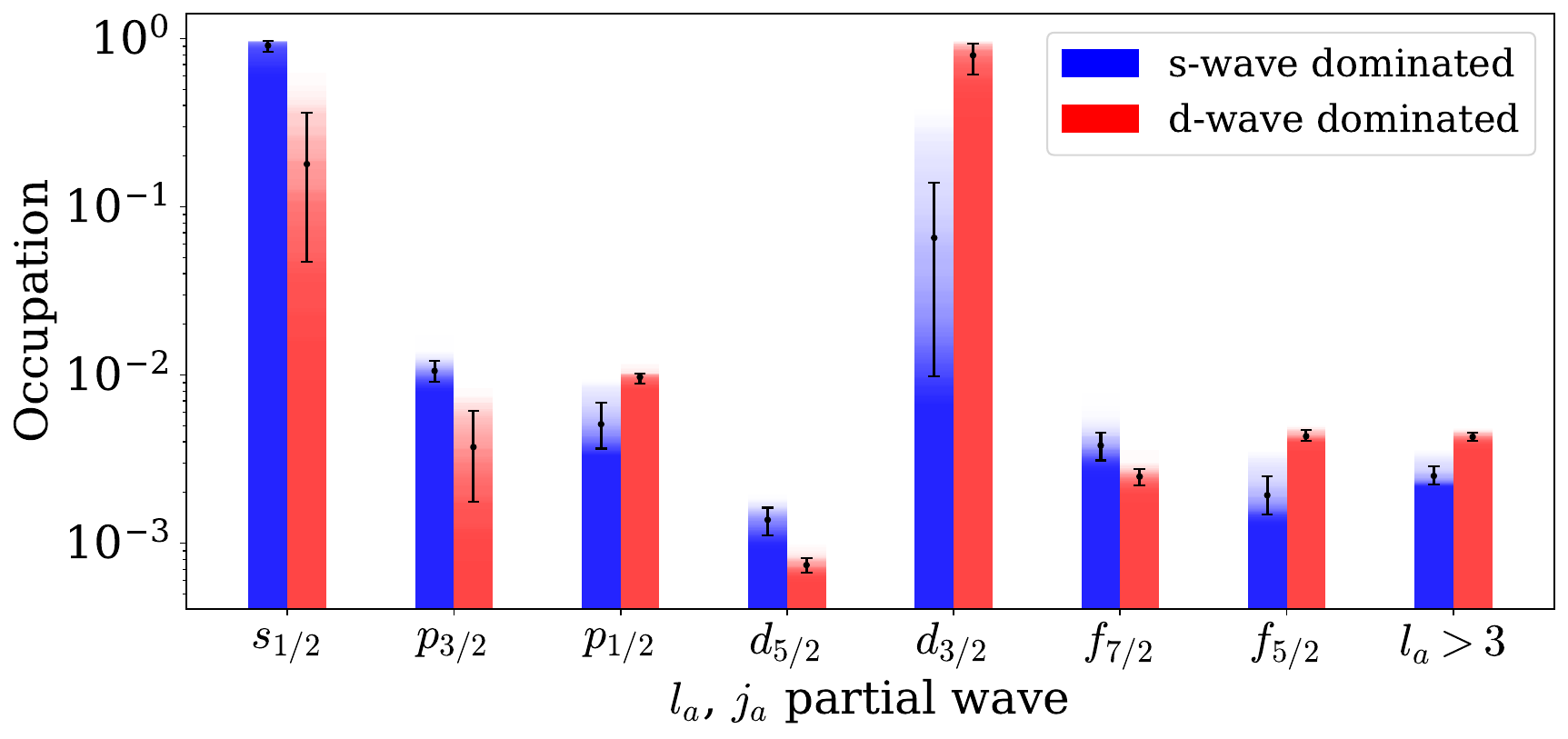}
	\caption{Occupation  number of $^{22}$C  ground state bound by 0.5 MeV  per partial waves of the core-neutron system  $l_a,j_a$. We separate the samples dominated by a $s_{1/2}$  (blue) and $d_{3/2}$ (red) configurations. The  varying opacity of the bar  correspond to the distribution of occupation numbers,  while points with error bars indicate the mean occupation and 68\% credible intervals. 
    }
	\label{fig:occupationsYbasis}
\end{figure}

\begin{figure}[t]
	\centering
	\includegraphics[width=0.8\linewidth]{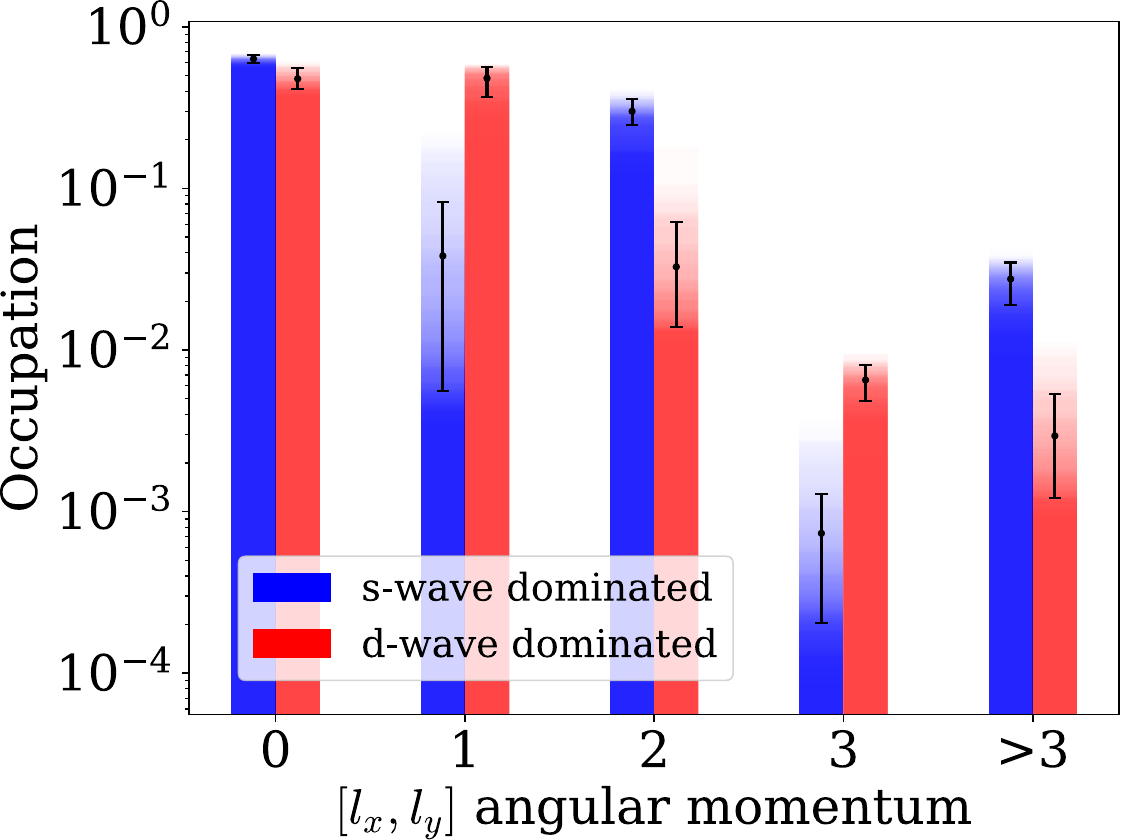}
	\caption{Same as Fig.~\ref{fig:occupationsYbasis}  but for $l_x$ the neutron-neutron angular momentum and $l_y$ the core-dineutron angular momentum, with $l_x=l_y$.
    }
	\label{fig:occupationsTbasis}
\end{figure}

We first analyze the partial-wave decomposition  of 
$^{22}$C ground states obtained in each group in the Y-basis in terms of $l_a$ and $j_a$ which couples the spin of one neutron to its orbital angular momentum $l_a$. Fig.~\ref{fig:occupationsYbasis} shows the occupation for the group leading to $\rho_{RMS}>6.5$~fm (blue histograms)  and $\rho_{RMS}<6.5$~fm (red histograms). The first group consists of  calculations which are almost entirely $(s_{1/2})^2$ with only minor contributions from  other partial waves. The second group corresponds to $d$-wave dominated calculations which are characterized by $(d_{3/2})^2$ configurations with  admixtures ranging from $4$ and $40$\% $s$-wave. No samples have significant contributions from any odd parities or from $d_{5/2}$, which is unsurprising since the  bound states in both the $d_{5/2}$ and  $p$-waves are Pauli-forbidden. Such different occupation motivates the name "$s$-wave dominated"   and "$d$-wave dominated" samples for  calculations with $\rho_{RMS}>6.5$~fm  and $\rho_{RMS}<6.5$~fm, respectively.

A complementary partial-wave decomposition of the ground state can be performed in the T-basis in Fig.~\ref{fig:occupationsTbasis}, which is more natural for interpreting $^{22}$C as a dineutron coupled to a $^{20}$C core.  
 Regardless of whether the configurations are  $s$- or $d$-wave dominated in the Y-basis,  all samples exhibit at least $\sim$50\% of $l_x=0$, corresponding to an $s$-wave dineutron. The main difference is in the different admixture: the $d$-wave dominated samples tend to see roughly equal mixture of this configuration with the $p$-wave dineutron component, while the $s$-wave dominated samples  tend to exhibit $d$-wave dineutron spectroscopic factors around 30\%.  We also verified that the dependence of these partial-wave decompositions in both the Y- and T-basis on the binding energy is small.

\begin{figure}[t]
	\centering
	\includegraphics[width=\linewidth]{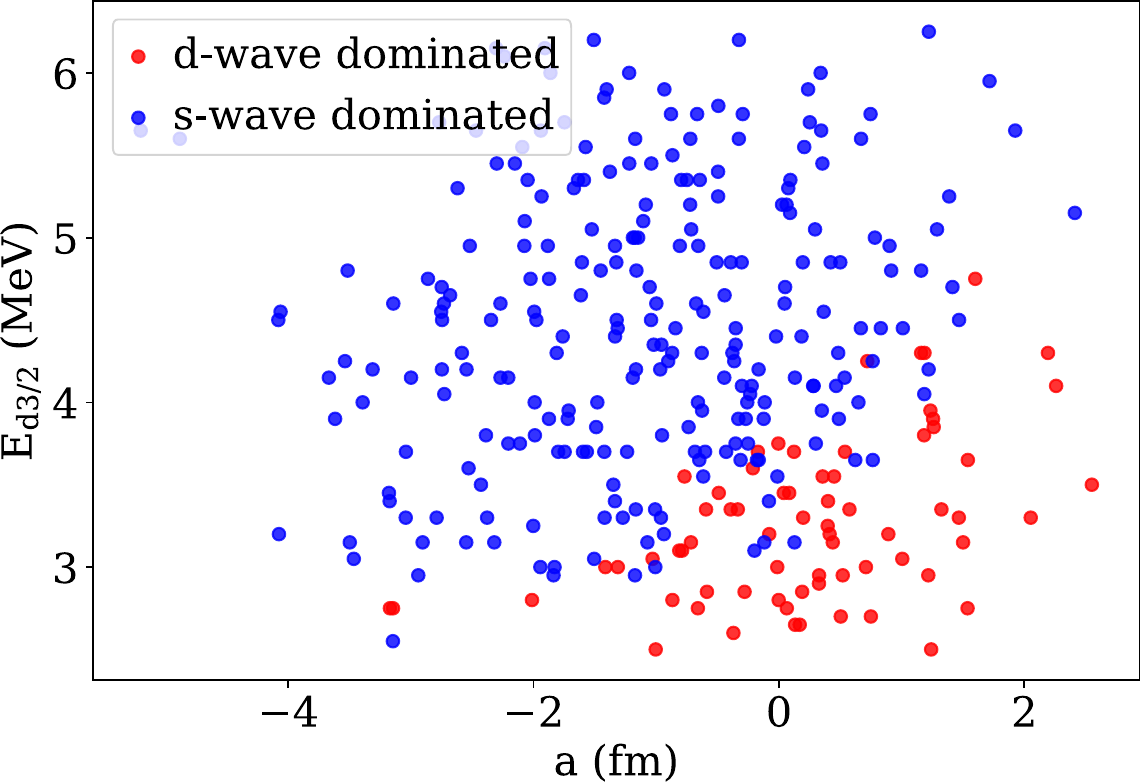}
	\caption{Scatter plot of  $d_{3/2}$ resonance energy and $s$-wave scattering length in $^{21}$C, coloured by whether the resultant three-body state has an rms hyperradius greater (resp. lower) than 6.5~fm, which is referred to as $s$- (resp. $d$-) wave dominated samples.}
	\label{fig:discrimination}
\end{figure}

Finally, we investigate how these two groups  relate to the properties of the $^{21}$C system. Fig.~\ref{fig:discrimination} shows all samples (blue and red points correspond  to $s$- and $d$-wave  configurations, respectively) as a function of the energy position of the $d_{3/2}$ resonance in $^{21}$C  and the $s$-wave scattering length. This figure shows that  $s$-wave (resp. $d$-wave) dominated samples are associated with  higher  (resp. lower) $d3/2$ resonance energies and  larger (resp. smaller) scattering lengths. This analysis confirms that observables such as rms radii and dipole strengths in the continuum are closely related to the partial-wave decomposition of the ground state of two-neutron halo nuclei, as well as to properties of the core-neutron system.

\section{Sensitivity of $dB(E1)/dE$ to scattering states and its uncertainty budget}
\label{sec:sensitivity}
\begin{figure}
	\centering
	\includegraphics[width=\linewidth]{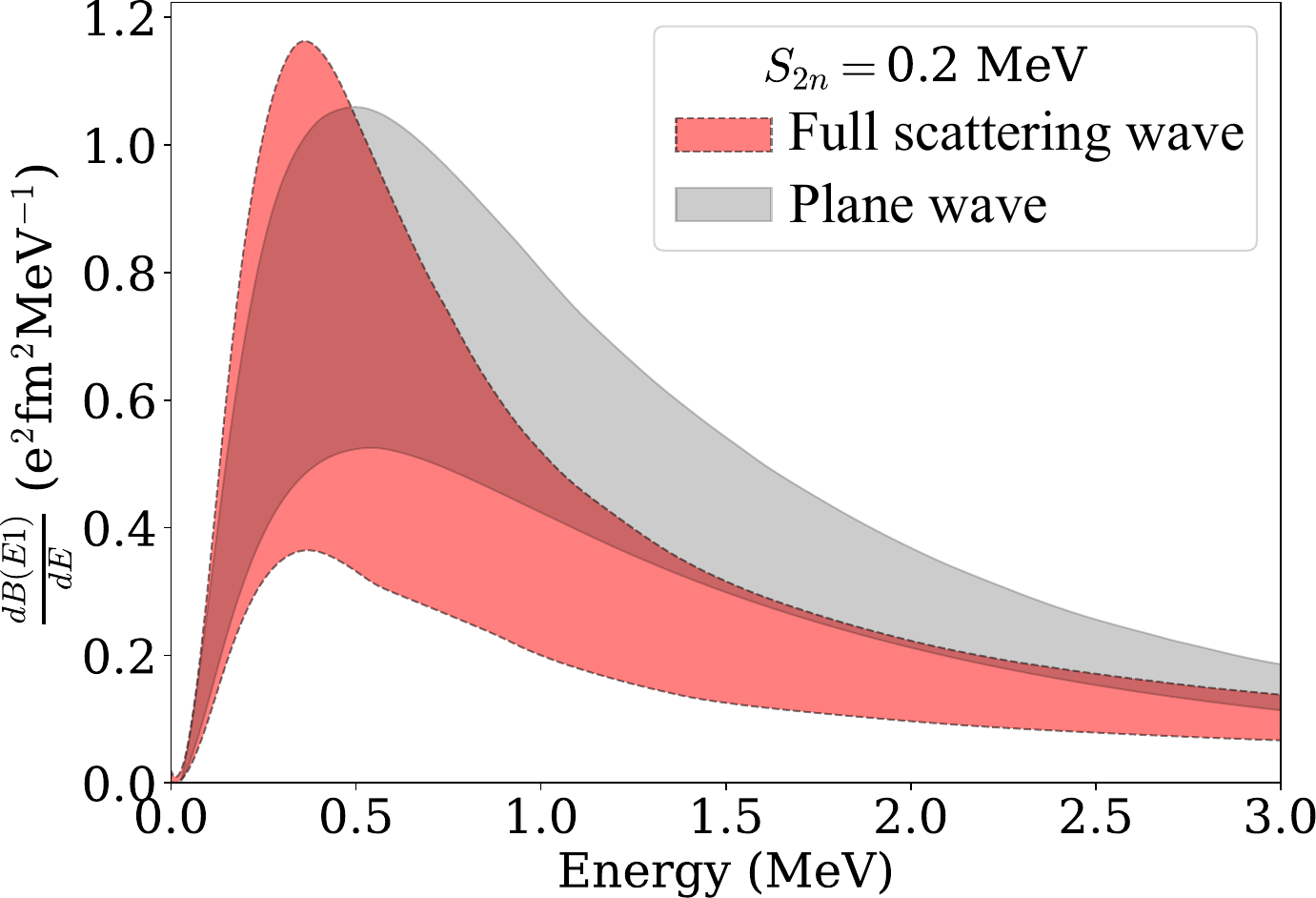}
	\caption{68\% credible intervals for the $dB(E1)/dE$ strength function calculated with $S_{2n}=0.2$~MeV using a plane wave approximation (gray band) and full scattering waves (red band, also shown in Fig.~\ref{fig:UQ}).}
	\label{fig:finalstateinteraction}
\end{figure}
In this appendix, we investigate the sensitivity of $dB(E1)/dE$ to the description of the three-body scattering states. First, we analyze the importance of FSI  in Fig.~\ref{fig:finalstateinteraction} by comparing dipole strength for a ground state bound by 0.2 MeV that are obtained  with the calculated three-body scattering waves along with similar calculations using plane waves, i.e., by using zero $^{20}$C-$n$ and $n$-$n$ potential to generate the scattering states.  As already mentioned in Sec.~\ref{22Cobservables}, FSI influence both the position of the peak and the  shape of the distribution. This indicates that   predictions  for $dB(E1)/dE$ using plane waves are not accurate.
Interestingly, the credible intervals are similar in both cases, indicating that most uncertainties in the dipole strength arises from the description of the ground state while uncertainties in the scattering states are comparatively smaller.

\begin{figure}
	\centering
	\includegraphics[width=\linewidth]{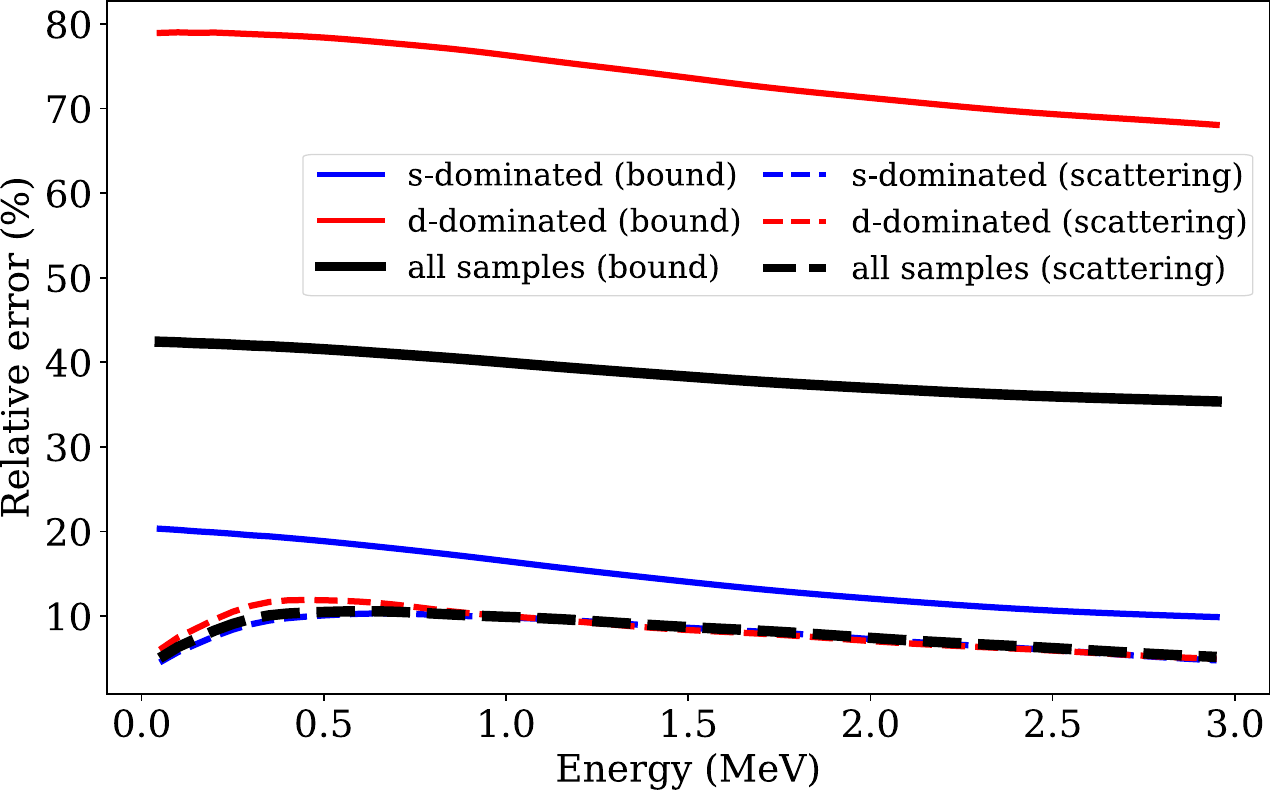}
	\caption{Relative error, defined as the ratio of the standard deviation of the $dB(E1)/dE$ sample to their mean, for calculations  varying only bound (solid lines) or scattering states (dashed lines). We consider all 315 samples (black lines), only the $s$-wave dominated  (blue lines) and only the $d$-wave dominates ones (red lines). }
	\label{fig:uncertainty}
\end{figure}

We further investigate the uncertainty budget of $dB(E1)/dE$ (that includes FSI) in Fig.~\ref{fig:uncertainty}, which shows the relative contribution of bound and scattering states to the total error.
The uncertainties associated with  the description of $^{22}$C ground state dominate the total uncertainty,  accounting for $\sim$40\% (solid black line) out of approximately a 50\% total error (see Fig.~\ref{fig:UQ}), while those associated with the scattering states are significantly smaller, around 10\% (dashed black line).
To assess the influence of the single-particle structure on the uncertainty budget of  $dB(E1)/dE$, we perform a similar analysis separately for samples which have a $s$- and $d$-wave dominated ground states (blue and red lines in Fig.~\ref{fig:uncertainty}). In both cases, the error on  the $dB(E1)/dE$ is  dominated by the ground-state description, while contribution from the scattering-sate uncertainties remains small.

\begin{figure}
	\centering
	\includegraphics[width=\linewidth]{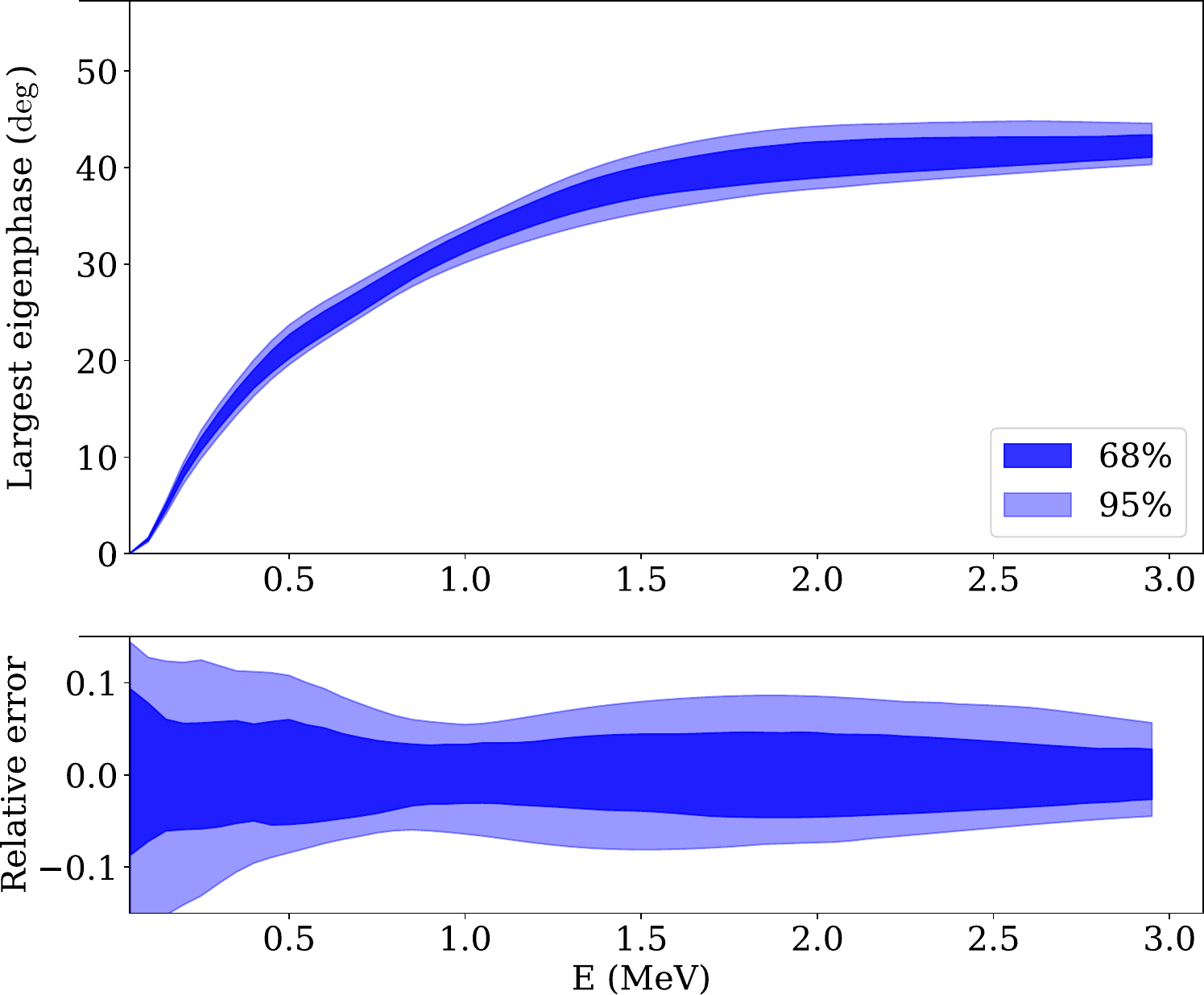}
	\caption{(Top) 68 and 95\% credible intervals for the largest eigenphase of the three-body $J^\pi=1^-$ scattering wave. (Bottom) relative error of the same.}
	\label{fig:phaseshift}
\end{figure}

For completeness, we provide in  Fig.~\ref{fig:phaseshift}  the 68\% and 95\% credible intervals for the largest eigen-phaseshift of the $1^- $ scattering waves (top panel) and the relative sizes of these intervals with respect to the mean (bottom panel). The relative error associated with  68\% credible interval is indeed relatively small, i.e., around 5\% across all energies.

\bibliographystyle{elsarticle-num} 
\bibliography{references}

@article{Ozawa2001,
    title = {Measurements of interaction cross sections for light neutron-rich nuclei at relativistic energies and determination of effective matter radii},
    volume = {691},
    issn = {0375-9474},
    url = {https://www.sciencedirect.com/science/article/pii/S0375947401005632},
    doi = {10.1016/S0375-9474(01)00563-2},
    journal = {Nucl. Phys. A},
    author = {Ozawa, A. and Bochkarev, O. and Chulkov, L. and C others},
    month = aug,
    year = {2001},
    pages = {599},
}

@article{Bertulani2002,
  title = {Effective Field Theory for Halo Nuclei: Shallow $p$-Wave States},
  shorttitle = {Effective Field Theory for Halo Nuclei},
  author = {Bertulani, C. A. and Hammer, H. -W. and {van Kolck}, U.},
  year = 2002,
  month = dec,
  journal = {Nucl. Phys. A},
  volume = {712},
  number = {1},
  pages = {37--58},
  issn = {0375-9474},
  doi = {10.1016/S0375-9474(02)01270-8},
  urldate = {2026-05-27},
  url={https://www.sciencedirect.com/science/article/pii/S0375947402012708?via%3Dihub}
}

@article{PhysRevLett.111.132501,
  title = {Efimov Physics Around the Neutron-Rich $^{60}\mathrm{Ca}$ Isotope},
  author = {Hagen, G. and Hagen, P. and Hammer, H.-W. and Platter, L.},
  journal = {Phys. Rev. Lett.},
  volume = {111},
  issue = {13},
  pages = {132501},
  numpages = {5},
  year = {2013},
  month = {Sep},
  publisher = {American Physical Society},
  doi = {10.1103/PhysRevLett.111.132501},
  url = {https://link.aps.org/doi/10.1103/PhysRevLett.111.132501}
}

@article{SANCSMHalo,
    author = {M. Huang and T. Frederico and P. Yin and R. A. M. Basili and P. J. Fasano and J. P. Vary},
    title = {Halo structure of 6He from ab initio two-nucleon spatial correlations},
    journal ={arXiv:2512.24123} ,
    year = 2025,
    url={https://arxiv.org/abs/2512.24123}
}

@article{Bonaiti2022,
  title = {Ab Initio Coupled-Cluster Calculations of Ground and Dipole Excited States in $^8${{He}}},
  author = {Bonaiti, F. and Bacca, S. and Hagen, G.},
  year = 2022,
  month = mar,
  journal = {Phys. Rev. C},
  volume = {105},
  number = {3},
  pages = {034313},
  publisher = {American Physical Society},
  url={https://journals.aps.org/prc/abstract/10.1103/PhysRevC.105.034313},
  doi = {10.1103/PhysRevC.105.034313},
  urldate = {2026-05-27}
}

@article{Hagen2010b,
  title = {Ab {{Initio Computation}} of the ${17}${{F}} {{Proton Halo State}} and {{Resonances}} in ${{A}}=17$ {{Nuclei}}},
  author = {Hagen, G. and Papenbrock, T. and {Hjorth-Jensen}, M.},
  year = 2010,
  month = may,
  url={https://journals.aps.org/prl/abstract/10.1103/PhysRevLett.104.182501},
  journal = {Phys. Rev. Lett.},
  volume = {104},
  number = {18},
  pages = {182501},
  publisher = {American Physical Society},
  doi = {10.1103/PhysRevLett.104.182501},
  urldate = {2026-05-27}
}

@article{Huang2026,
  title = {Halo Structure of 6 {{He}} from Ab Initio Two-Nucleon Spatial Correlations},
  author = {{Anonymous}},
  year = 2026,
  month = may,
  journal = {Phys. Rev. C},
  issn = {2469-9985, 2469-9993},
  doi = {10.1103/hy1r-4vzg},
  urldate = {2026-06-01},
  langid = {english}
}

@article{Hagen2013a,
  title = {Efimov {{Physics Around}} the {{Neutron-Rich}} \$\textasciicircum\textbraceleft 60\textbraceright\textbackslash mathrm\textbraceleft{{Ca}}\textbraceright\$ {{Isotope}}},
  author = {Hagen, G. and Hagen, P. and Hammer, H.-W. and Platter, L.},
  year = 2013,
  month = sep,
  journal = {Phys. Rev. Lett.},
  volume = {111},
  number = {13},
  pages = {132501},
  publisher = {American Physical Society},
  doi = {10.1103/PhysRevLett.111.132501},
  urldate = {2026-06-01}
}

@article{Hiyama2019,
  title = {Modeling {{B}} 19 as a {{B}} 17 - n - n Three-Body System in the Unitary Limit},
  author = {Hiyama, Emiko and Lazauskas, Rimantas and Marqu{\'e}s, F. Miguel and Carbonell, Jaume},
  year = 2019,
  month = jul,
  journal = {Phys. Rev. C},
  volume = {100},
  number = {1},
  pages = {011603},
  issn = {2469-9985, 2469-9993},
  doi = {10.1103/PhysRevC.100.011603},
  urldate = {2026-06-01},
  langid = {english}
}

@article{Hiyama2022,
  title = {Scaling of the {{B}} 19 Two-Neutron Halo Properties Close to Unitarity},
  author = {Hiyama, Emiko and Lazauskas, Rimantas and Carbonell, Jaume and Frederico, Tobias},
  year = 2022,
  month = dec,
  journal = {Phys. Rev. C},
  volume = {106},
  number = {6},
  pages = {064001},
  issn = {2469-9985, 2469-9993},
  doi = {10.1103/PhysRevC.106.064001},
  urldate = {2026-06-01},
  langid = {english}
}

@article{Hagen2013,
  title = {Charge Form Factors of Two-Neutron Halo Nuclei in Halo {{EFT}}},
  author = {Hagen, P. and Hammer, H. -W. and Platter, L.},
  year = 2013,
  month = sep,
  journal = {Eur. Phys. J. A},
  volume = {49},
  number = {9},
  pages = {118},
  issn = {1434-601X},
  doi = {10.1140/epja/i2013-13118-4},
  url={https://link.springer.com/article/10.1140/epja/i2013-13118-4},
  urldate = {2026-05-27},
  langid = {english}
}

@article{Hammer2020,
  title = {Nuclear Effective Field Theory: {{Status}} and Perspectives},
  shorttitle = {Nuclear Effective Field Theory},
  author = {Hammer, H.-W. and K{\"o}nig, Sebastian and {van Kolck}, U.},
  year = 2020,
  month = jun,
  journal = {Rev. Mod. Phys.},
  volume = {92},
  number = {2},
  pages = {025004},
  publisher = {American Physical Society},
  doi = {10.1103/RevModPhys.92.025004},
  urldate = {2026-05-27},
url={https://journals.aps.org/rmp/abstract/10.1103/RevModPhys.92.025004}}

@article{Braun17C, 
author={J. Braun and H.-W. Hammer and L. Platter},
title={Halo structure of $^{17}${{C}}},
journal={Eur. Phys. J. A},
volume={54}, 
issue={196},

year=2018,
url={ http://dx.doi.org/10.1140/epja/i2018-12630-3},
doi={10.1140/epja/i2018-12630-3}}

@article{Li2024b,
  title = {Unveiling Potential Neutron Halos in Intermediate-Mass Nuclei: {{An}} {\emph{Ab Initio}} Study},
  shorttitle = {Unveiling Potential Neutron Halos in Intermediate-Mass Nuclei},
  author = {Li, H. H.},
  year = 2024,
  journal = {Phys. Rev. C},
  volume = {109},
  number = {6},
  url={https://journals.aps.org/prc/abstract/10.1103/PhysRevC.109.L061304},
  doi = {10.1103/PhysRevC.109.L061304}
}

@article{Rodkin2021,
  title = {Detailed Theoretical Study of the Decay Properties of States in the $^7${{He}} Nucleus within an Ab Initio Approach},
  author = {Rodkin, D. M. and Tchuvil'sky, {\relax Yu}. M.},
  url={https://journals.aps.org/prc/abstract/10.1103/PhysRevC.104.044323},
  year = 2021,
  month = oct,
  journal = {Phys. Rev. C},
  volume = {104},
  number = {4},
  pages = {044323},
  publisher = {American Physical Society},
  doi = {10.1103/PhysRevC.104.044323},
  urldate = {2026-05-27}
}

@article{Shen2026,
  title = {Ab {{Initio Study}} on the {{Halo Structure}} in $^{11}${{Be}}},
  author = {Shen, Shihang and Elhatisari, Serdar and Lee, Dean and Mei{\ss}ner, Ulf-G. and Ren, Zhengxue},
  year = 2026,
  month = mar,
  journal = {Particles},
  volume = {9},
  number = {1},
  pages = {25},
  publisher = {Multidisciplinary Digital Publishing Institute},
  issn = {2571-712X},
  url={https://www.mdpi.com/2571-712X/9/1/25},
  doi = {10.3390/particles9010025},
  urldate = {2026-05-27},
  copyright = {http://creativecommons.org/licenses/by/3.0/},
  langid = {english}
}

@article{emcee,
    title={emcee: The {{MCMC}} {{H}}ammer},
    volume={125},
    journal={Publ. Atron. Soc. Pac.},
    year={2013},
    author={Foreman-Mackey, D. and Hogg, D. W. and others},
    doi={10.1086/670067}
}

@phdthesis{Leblond2015,
    type = {Theses},
    title = {Structure des isotopes de bore et de carbone riches en neutrons aux limites de la stabilité},
    url = {https://theses.hal.science/tel-01289381},
    school = {Normandie Université, France},
    author = {Leblond, Sylvain},    
    year = {2015}
}

@article{cxcz-z8px,
  title = {Comparing invariant-mass spectroscopy of $^{8}\mathrm{B}$ with ab initio predictions},
  author = {Charity, R. J. and Sargsyan, G. H. and Launey, K. D. and Webb, T.B. and Brown, K.W. and Sobotka, L. G.},
  journal = {Phys. Rev. C},
  volume = {113},
  issue = {2},
  pages = {024322},
  numpages = {15},
  year = {2026},
  month = {Feb},
  publisher = {American Physical Society},
  doi = {10.1103/cxcz-z8px},
  url = {https://link.aps.org/doi/10.1103/cxcz-z8px}
}

@article{Mosby2013,
    title = {Search for $^{21}${C} and constraints on $^{22}${C}},
    volume = {909},
    issn = {0375-9474},
    url = {https://www.sciencedirect.com/science/article/pii/S0375947413004843},
    doi = {10.1016/j.nuclphysa.2013.04.004},
    journal = {Nucl. Phys. A},
    author = {Mosby, S. and Badger, N. S. and Baumann, T. and others},    
    year = {2013},
    pages = {69},
}

@article{Pinilla2016,
    title = {Coulomb breakup of $^{22}${C}  in a four-body model},
    volume = {94},
    url = {https://link.aps.org/doi/10.1103/PhysRevC.94.024620},
    doi = {10.1103/PhysRevC.94.024620},
    language = {en},    
    urldate = {2024-02-28},
    journal = {Phys. Rev. C},
    author = {Pinilla, E. C. and Descouvemont, P.},
    year = {2016},
    pages = {024620},
}

@article{Thompson1977,
    title = {Systematic investigation of scattering problems with the resonating-group method},
    volume = {286},
    issn = {0375-9474},
    url = {https://www.sciencedirect.com/science/article/pii/0375947477900070},
    doi = {10.1016/0375-9474(77)90007-0},    
    journal = {Nucl. Phys. A},
    author = {Thompson, D. R. and Lemere, M. and Tang, Y. C.},    
    year = {1977},
    pages = {53},
}

@article{Varga1995,
    title = {Precise solution of few-body problems with the stochastic variational method on a correlated {Gaussian} basis},
    volume = {52},
    issn = {0556-2813, 1089-490X},
    url = {https://link.aps.org/doi/10.1103/PhysRevC.52.2885},
    doi = {10.1103/PhysRevC.52.2885},
    journal = {Phys. Rev. C},
    author = {Varga, K. and Suzuki, Y.},
    year = {1995},
    pages = {2885},
}

@article{Bogner2011,
    title = {Testing the density matrix expansion against \textit{ab initio} calculations of trapped neutron drops},
    volume = {84},
    copyright = {http://link.aps.org/licenses/aps-default-license},
    issn = {0556-2813, 1089-490X},
    url = {https://link.aps.org/doi/10.1103/PhysRevC.84.044306},
    doi = {10.1103/PhysRevC.84.044306},
    journal = {Phys. Rev. C},
    author = {Bogner, S. K. and Furnstahl, R. J. and Hergert, H. and others},    
    year = {2011},
    pages = {044306},
}

@article{Acharya2013,
    title = {Implications of a matter-radius measurement for the structure of {Carbon}-22},
    volume = {723},
    issn = {03702693},
    url = {http://arxiv.org/abs/1303.6720},
    doi = {10.1016/j.physletb.2013.04.055},
    journal = {Phys. Lett. B},
    author = {Acharya, B. and Ji, C. and Phillips, D. R.},    
    year = {2013},
    pages = {196},
}

@article{Yamashita2011,
    title = {Constraints on two-neutron separation energy in the {Borromean} $^{22}${C} nucleus},
    volume = {697},
    issn = {0370-2693},
    url = {https://www.sciencedirect.com/science/article/pii/S0370269311000773},
    doi = {10.1016/j.physletb.2011.01.040},
    journal = {Phys. Lett. B},
    author = {Yamashita, M. T. and de Carvalho, R. S. Marques and Frederico, T. and  others},
    year = {2011},
    pages = {90},
}

@article{Horiuchi2006,
    title = {$^{22}${C} : {An} $s$-wave two-neutron halo nucleus},
    volume = {74},
    url = {https://link.aps.org/doi/10.1103/PhysRevC.74.034311},
    doi = {10.1103/PhysRevC.74.034311},
    journal = {Phys. Rev. C},
    author = {Horiuchi, W. and Suzuki, Y.},
    month = sep,
    year = {2006},
    pages = {034311},
}

@article{Hongo2022,
    title = {Universal Properties of Weakly Bound Two-Neutron Halo Nuclei},
    volume = {128},
    url = {https://link.aps.org/doi/10.1103/PhysRevLett.128.212501},
    doi = {10.1103/PhysRevLett.128.212501},
    journal = {Phys. Rev. Lett.},
    author = {Hongo, Masaru and Son, Dam Thanh},
    year = {2022},
    pages = {212501}
}

@article{Tanaka2010,
    title = {Observation of a Large Reaction Cross Section in the Drip-Line Nucleus $^{22}${C}},
    volume = {104},
    issn = {0031-9007, 1079-7114},
    url = {https://link.aps.org/doi/10.1103/PhysRevLett.104.062701},
    doi = {10.1103/PhysRevLett.104.062701},
    journal = {Phys. Rev. Lett.},
    author = {Tanaka, K. and Yamaguchi, T. and Suzuki, T. and others},    
    year = {2010},
    pages = {062701},
}

@article{Togano2016,
    title = {Interaction cross section study of the two-neutron halo nucleus $^{22}${C}},
    volume = {761},   
    url = {https://www.sciencedirect.com/science/article/pii/S0370269316304890},
    doi = {10.1016/j.physletb.2016.08.062}, 
    journal = {Phys. Lett. B},
    author = {Togano, Y. and Nakamura, T. and Kondo, Y. and others},    
    year = {2016},
    pages = {412},
}

@article{Cook2020,
    title = {Halo Structure of the Neutron-Dripline Nucleus $^{19}${B}},
    volume = {124},
    url = {https://link.aps.org/doi/10.1103/PhysRevLett.124.212503},
    doi = {10.1103/PhysRevLett.124.212503},
    journal = {Phys. Rev. Lett.},
    publisher = {American Physical Society},
    author = {Cook, K. J. and Nakamura, T. and Kondo, Y. and other},    
    year = {2020},
    pages = {212503}
}

@article{Aumann1999a,
    title = {Continuum excitations in $^6${He}},
    volume = {59},
    url = {https://link.aps.org/doi/10.1103/PhysRevC.59.1252},
    doi = {10.1103/PhysRevC.59.1252},
    urldate = {2026-03-10},
    journal = {Phys. Rev. C},
    publisher = {American Physical Society},
    author = {Aumann, T. and Aleksandrov, D. and Axelsson, L. and others},    
    year = {1999},
    pages = {1252},
}

@article{Wang2002a,
    title = {Dissociation of $^6${He}},
    volume = {65},
    url = {https://link.aps.org/doi/10.1103/PhysRevC.65.034306},
    doi = {10.1103/PhysRevC.65.034306},
    journal = {Phys. Rev. C},
    publisher = {American Physical Society},
    author = {Wang, J. and Galonsky, A. and Kruse, J. J. and others},
    year = {2002},
    pages = {034306}
}

@article{Nakamura2006a,
    title = {Observation of Strong Low-Lying ${E1}$ Strength in the Two-Neutron Halo Nucleus $^{11}${Li}},
    volume = {96},
    url = {https://link.aps.org/doi/10.1103/PhysRevLett.96.252502},
    doi = {10.1103/PhysRevLett.96.252502},
    journal = {Phys. Rev. Lett.},
    publisher = {American Physical Society},
    author = {Nakamura, T. and Vinodkumar, A. M. and Sugimoto, T. and others},    
    year = {2006},
    pages = {252502},
}

@article{Aumann2013,
    title = {The electric dipole response of exotic nuclei},
    volume = {T152},
    issn = {0031-8949, 1402-4896},
    url = {https://iopscience.iop.org/article/10.1088/0031-8949/2013/T152/014012},
    doi = {10.1088/0031-8949/2013/T152/014012},
    journal = {Phys. Script.},
    author = {Aumann, T and Nakamura, T},    
    year = {2013},
    pages = {014012},
}

@article{Gobel2024,
    title = {Universality of $nn$ distributions of $s$-wave $2n$ halo nuclei and the unitary limit},
    volume = {110},
    url = {https://link.aps.org/doi/10.1103/PhysRevC.110.024003},
    doi = {10.1103/PhysRevC.110.024003},
    journal = {Phys. Rev. C},
    publisher = {American Physical Society},
    author = {Göbel, Matthias and Hammer, Hans-Werner and Phillips, Daniel R.},    
    year = {2024},
    pages = {024003},
}

@article{Hammer2017a,
    title = {Effective field theory description of halo nuclei},   
    author = {Hammer, H-W and Phillips, D. R. and Ji, Chen},
    year = {2017},
    journal={J. Phys. G: Nucl. Part. Phys.},
    volume=44,
    pages=103002,
    url={https://iopscience.iop.org/article/10.1088/1361-6471/aa83db},
    doi={10.1088/1361-6471/aa83db}
}

@article{Gaudefroy2012,
    title = {Direct Mass Measurements of $^{19}${B}, $^{22}${C}, $^{29}${F}, $^{31}${Ne}, $^{34}${Na} and other light exotic nuclei},
    volume = {109},
    url = {https://link.aps.org/doi/10.1103/PhysRevLett.109.202503},
    doi = {10.1103/PhysRevLett.109.202503},
    journal = {Phys. Rev. Lett.},
    author = {Gaudefroy, L. and Mittig, W. and Orr, N. A. and others},    
    year = {2012},
    pages = {202503},
}

@article{Wang2021,
    title = {The {AME} 2020 atomic mass evaluation ({II}). {Tables}, graphs and references*},
    volume = {45},
    url = {https://doi.org/10.1088/1674-1137/abddaf},
    doi = {10.1088/1674-1137/abddaf},
    journal = {Chin. Phys. C},
    author = {Wang, Meng and Huang, W.J. and Kondev, F.G. and others},
    year = {2021},
    pages = {030003},
}

@misc{McGlynn2026,
    title = {Development of an accurate formalism to predict properties of two-neutron halo nuclei: case study of $^{22}${C}},
    url = {http://arxiv.org/abs/2602.15765},
    publisher = {arXiv},
    author = {McGlynn, Patrick and Hebborn, Chloë},
    year = {2026},
    note = {arXiv:2602.15765 [nucl-th]}
}

@article{Bazin2023,
    title = {Perspectives on {Few}-{Body} {Cluster} {Structures} in {Exotic} {Nuclei}},
    volume = {64},
    issn = {1432-5411},
    url = {https://doi.org/10.1007/s00601-023-01794-0},
    doi = {10.1007/s00601-023-01794-0},
    journal = {Few-Body Syst.},
    author = {Bazin, Daniel and Becker, Kevin and Bonaiti, Francesca and others},  
    year = {2023},
    pages = {25},
}

@article{Freer2018,
    title = {Microscopic clustering in light nuclei},
    volume = {90},
    url = {https://link.aps.org/doi/10.1103/RevModPhys.90.035004},
    doi = {10.1103/RevModPhys.90.035004},
    journal = {Rev. Mod. Phys},
    author = {Freer, Martin and Horiuchi, Hisashi and Kanada-En’yo, Yoshiko and others},    
    year = {2018},
    pages = {035004},
}

@article{Tanihata2013,
    title = {Recent experimental progress in nuclear halo structure studies},
    volume = {68},
    issn = {0146-6410},
    url = {https://www.sciencedirect.com/science/article/pii/S0146641012001081},
    doi = {10.1016/j.ppnp.2012.07.001},
    journal = {Prog. Part. Nucl. Phys.},
    author = {Tanihata, Isao and Savajols, Herve and Kanungo, Rituparna},    
    year = {2013},
    pages = {215},
}

@article{Casal2019a,
    title = {Identifying structures in the continuum: {Application} to $^{16}${Be}},
    volume = {99},
    shorttitle = {Identifying structures in the continuum},
    url = {https://link.aps.org/doi/10.1103/PhysRevC.99.014604},
    doi = {10.1103/PhysRevC.99.014604},   
    journal = {Phys. Rev. C},
    publisher = {American Physical Society},
    author = {Casal, J. and Gómez-Camacho, J.},    
    year = {2019},
    pages = {014604},
}

@article{Lovell2017,
    title = {Three-body model for the two-neutron emission of $^{16}${Be}},
    volume = {95},
    copyright = {http://link.aps.org/licenses/aps-default-license},
    issn = {2469-9985, 2469-9993},
    url = {https://link.aps.org/doi/10.1103/PhysRevC.95.034605},
    doi = {10.1103/PhysRevC.95.034605},
    journal = {Phys. Rev. C},
    author = {Lovell, A. E. and Nunes, F. M. and Thompson, I. J.},    
    year = {2017},
    pages = {034605},
}

@article{Costa2025,
    title = {Effective field theory for weakly bound two-neutron halo nuclei: {Corrections} from neutron-neutron effective range},
    volume = {112},
    shorttitle = {Effective field theory for weakly bound two-neutron halo nuclei},
    url = {https://link.aps.org/doi/10.1103/lds3-g3tp},
    doi = {10.1103/lds3-g3tp},
    journal = {Phys. Rev. C},
    publisher = {American Physical Society},
    author = {Costa, Davi B. and Hongo, Masaru and Son, Dam Thanh},    
    year = {2025},
    pages = {014001},
}

@article{Monteagudo2024,
    title = {Mass, {Spectroscopy}, and {Two}-{Neutron} {Decay} of  $^{16}${Be}},
    volume = {132},
    url = {https://link.aps.org/doi/10.1103/PhysRevLett.132.082501},
    doi = {10.1103/PhysRevLett.132.082501},  
    journal = {Phys. Rev. Lett.},
    publisher = {American Physical Society},
    author = {Monteagudo, B. and Marqués, F. M. and Gibelin, J. and others},    
    year = {2024},
    pages = {082501},
}

@article{Casal2020a,
    title = {Three-body structure of $^{19}${B}: {Finite}-range effects in two-neutron halo nuclei},
    volume = {102},

    url = {https://link.aps.org/doi/10.1103/PhysRevC.102.051304},
    doi = {10.1103/PhysRevC.102.051304},
    journal = {Phys. Rev. C},
    author = {Casal, J. and Garrido, E.},    
    year = {2020},
    pages = {051304},
}

@article{Ershov2012,
    title = {Binding energy constraint on matter radius and soft dipole excitations of $^{22}${C}},
    volume = {86},       
    journal = {Phys. Rev. C},
    author = {Ershov, S. N.},
    year = {2012},
    url={https://link.aps.org/doi/10.1103/PhysRevC.86.034331},
    doi={10.1103/PhysRevC.86.034331}
}

@article{Fortunato2020,
    title = {The $^{29}${F} nucleus as a lighthouse on the coast of the island of inversion},
    volume = {3},
    url = {https://www.nature.com/articles/s42005-020-00402-5},
    doi = {10.1038/s42005-020-00402-5},
    journal = {Commun. Phys.},
    author = {Fortunato, L. and Casal, J. and Horiuchi, W. and others},
    year = {2020},
    pages = {132},
}

@article{Singh2024a,
    title = {Prediction of two-neutron halos in the \textit{{N}}=28 isotones $^{40}${Mg} and $^{39}${Na}},
    volume = {853},
    url = {https://www.sciencedirect.com/science/article/pii/S0370269324002521},
    doi = {10.1016/j.physletb.2024.138694},
    urldate = {2026-02-09},
    journal = {Phys. Lett. B},
    author = {Singh, Jagjit and Casal, J. and Horiuchi, W. and  others},
    year = {2024},
    pages = {138694},
}

@article{Poves2017,
    title = {Shell model spectroscopy far from stability},
    volume = {44},
    issn = {13616471},
    doi = {10.1088/1361-6471/aa7789},
    
    number = {8},
    journal = {J. Phys. G: Nucl. Part. Phys.},
    publisher = {Institute of Physics Publishing},
    author = {Poves, A.},
    url={https://iopscience.iop.org/article/10.1088/1361-6471/aa7789},
    year = {2017},
}

@article{Nunes1996a,
    title = {Core excitation in three-body systems: {Application} to $^{12}${Be}},
    volume = {609},
    url = {https://www.sciencedirect.com/science/article/pii/0375947496002849},
    doi = {10.1016/0375-9474(96)00284-9},
    journal = {Nucl. Phys. A},
    author = {Nunes, F. M. and Christley, J. A. and Thompson, I. J. and  others},
    year = {1996},
    pages = {43},
}

@article{Tostevin2001,
    title = {Calculations of three-body observables in $^8${B} breakup},
    volume = {63},
    url = {https://link.aps.org/doi/10.1103/PhysRevC.63.024617},
    doi = {10.1103/PhysRevC.63.024617},
    urldate = {2026-02-13},
    journal = {Phys. Rev. C},
    publisher = {American Physical Society},
    author = {Tostevin, J. A. and Nunes, F. M. and Thompson, I. J.},
    year = {2001},
    pages = {024617}
}

@article{Thompson2004c,
    title = {{FaCE}: a tool for three body {Faddeev} calculations with core excitation},
    volume = {161},
    url = {https://www.sciencedirect.com/science/article/pii/S0010465504002140},
    doi = {10.1016/j.cpc.2004.03.007},
    journal = {Comp. Phys. Commun.},
    author = {Thompson, I. J. and Nunes, F. M. and Danilin, B. V.},    
    year = {2004},
    pages = {87},
}

@article{Pinilla2025,
    title = {Three-body model of $^6${He} with nonlocal halo effective field theory potentials},
    volume = {112},
    url = {https://link.aps.org/doi/10.1103/kskt-7p8g},
    doi = {10.1103/kskt-7p8g},
    journal = {Phys. Rev. C},
    publisher = {American Physical Society},
    author = {Pinilla, E. C. and Leidemann, W. and Orlandini, G. and other},
    year = {2025},
    pages = {024003},
}

@article{Descouvemont2003,
    title = {Three-body systems with {Lagrange}-mesh techniques in hyperspherical coordinates},
    volume = {67},
    url = {https://link.aps.org/doi/10.1103/PhysRevC.67.044309},
    doi = {10.1103/PhysRevC.67.044309},    
    journal = {Phys. Rev. C},
    author = {Descouvemont, P. and Daniel, C. and Baye, D.},
    month = apr,
    year = {2003},
    pages = {044309},
}

@article{Descouvemont2006,
    title = {Three-body continuum states on a {Lagrange} mesh},
    volume = {765},
    issn = {0375-9474},
    url = {https://www.sciencedirect.com/science/article/pii/S037594740501198X},
    doi = {10.1016/j.nuclphysa.2005.11.010},
    journal = {Nucl. Phys. A},
    author = {Descouvemont, P. and Tursunov, E. and Baye, D.},
    year = {2006},
    pages = {370},
}

@article{Labiche2001,
    title = {Halo {Structure} of $^{14}${Be}},
    volume = {86},
    shorttitle = {Halo {Structure} of {\textless}span class="aps-inline-formula"{\textgreater}{\textless}math xmlns="http},
    doi = {10.1103/PhysRevLett.86.600},
    journal = {Phys. Rev. Lett.},
    author = {Labiche, M.},
    year = {2001},
    pages = {600},
    url={https://journals.aps.org/prl/abstract/10.1103/PhysRevLett.86.600}
}

@article{Kobayashi2012,
    title = {One- and two-neutron removal reactions from the most neutron-rich carbon isotopes},
    volume = {86},
    url = {https://link.aps.org/doi/10.1103/PhysRevC.86.054604},
    doi = {10.1103/PhysRevC.86.054604},
    journal = {Phys. Rev. C},
    publisher = {American Physical Society},
    author = {Kobayashi, N. and Nakamura, T. and Tostevin, J. A. and others},    
    year = {2012},
    pages = {054604},
}

@article{Nagahisa2018,
    title = {Examination of the $^{22}${C}  radius determination with interaction cross sections},
    volume = {97},
    issn = {2469-9985, 2469-9993},
    url = {https://link.aps.org/doi/10.1103/PhysRevC.97.054614},
    doi = {10.1103/PhysRevC.97.054614},    
    journal = {Phys. Rev. C},
    author = {Nagahisa, T. and Horiuchi, W.},
    month = may,
    year = {2018},
    pages = {054614},
}

@article{Bagchi2020,
    title = {Two-Neutron Halo is Unveiled in $^{29}${F}},
    volume = {124},
    url = {https://link.aps.org/doi/10.1103/PhysRevLett.124.222504},
    doi = {10.1103/PhysRevLett.124.222504},    
    journal = {Phys. Rev. Lett.},
    publisher = {American Physical Society},
    author = {Bagchi, S. and Kanungo, R. and Tanaka, Y. K. and others},
    year = {2020},
    pages = {222504},
}

@article{Smith2026,
    author = {A. J. Smith and K. Godbey and C. Hebborn and others},
    title = {Matter radii from interaction cross sections using microscopic nuclear densities},
    journal = {arXiv:2603.18862},
    year =2026,
    url={https://arxiv.org/abs/2603.18862}
}

@article{Pruitt2024,
  title = {Role of the likelihood for elastic scattering uncertainty quantification},
  author = {Pruitt, C. D. and Lovell, A. E. and Hebborn, C. and others},
  journal = {Phys. Rev. C},
  volume = {110},
  issue = {6},
  pages = {064606},
  numpages = {9},
  year = {2024},
  publisher = {American Physical Society},
  doi = {10.1103/PhysRevC.110.064606},
  url = {https://link.aps.org/doi/10.1103/PhysRevC.110.064606}
}

@article{BANDmanifesto,
author={D. R. Phillips and R. J. Furnstahl and U. Heinz and others}, 
title={Get on the {BAND} Wagon: a {B}ayesian framework for quantifying model uncertainties in nuclear dynamics},
year=2021,
journal={J. Phys. G: Nucl. Part. Phys.},
volume=48,
pages=072001,
doi={10.1088/1361-6471/abf1df},
url={https://iopscience.iop.org/article/10.1088/1361-6471/abf1df}}

@article{navratil_ab_2011,
    title = {Ab initio many-body calculation of the $^{7}${{Be}}(p,$\gamma$)$^{8}${{B}}  radiative capture},
    volume = {704},
    issn = {0370-2693},
    url = {https://www.sciencedirect.com/science/article/pii/S0370269311011646},
    doi = {10.1016/j.physletb.2011.09.079},   
    journal = {Phys. Lett. B},
    author = {Navrátil, Petr and Roth, Robert and Quaglioni, Sofia},
    month = oct,
    year = {2011},
    pages = {379},
}

@article{reviewncsmc,
title={Unified ab initio approaches to nuclear structure and reactions},
author={Navratil, Petr and Quaglioni, Sofia and Hupin, Guillaume and others},
year=2016,
journal={Phys. Scr.},
volume=91,
pages=053002,
doi={10.1088/0031-8949/91/5/053002},
url={https://iopscience.iop.org/article/10.1088/0031-8949/91/5/053002}}

@article{Kravvaris8B,
    author = {K. Kravvaris and
 P. Navrátil and S. Quaglioni and others},
    title = {Ab initio informed evaluation of the radiative capture of protons on $^7${Be}},
    journal = {Phys. Lett. B},
    year = 2023,
    volumne=845,
    pages=138156,
    doi={10.1016/j.physletb.2023.138156},
    url={https://www.sciencedirect.com/science/article/pii/S0370269323004902?via%3Dihub}
}

@misc{navratil_halo_2026,
    title = {Halo {Nuclei} from {Ab} {Initio} {Nuclear} {Theory}},
    url = {http://arxiv.org/abs/2604.02612},
    doi = {10.48550/arXiv.2604.02612},
    urldate = {2026-04-06},
    publisher = {arXiv},
    author = {Navratil, Petr and Quaglioni, Sofia and Hupin, Guillaume and others},    
    year = {2026},
    note = {arXiv:2604.02612 [nucl-th]},
}

@article{quaglioni_three-cluster_2018,
    title = {Three-cluster dynamics within the \textit{ab initio} no-core shell model with continuum: {How} many-body correlations and $\alpha$-clustering shape $^{6}${He}},
    volume = {97},
    shorttitle = {Three-cluster dynamics within the \textit{ab initio} no-core shell model with continuum},
    doi = {10.1103/PhysRevC.97.034332},    
    journal = {Phys. Rev. C},
    author = {Quaglioni, Sofia and Romero-Redondo, Carolina and Navratil, Petr and others},
    year = {2018},
    url = {https://link.aps.org/doi/10.1103/PhysRevC.97.034332},
}

@article{Letter6HeNCSMC,
  title = {How Many-Body Correlations and $\ensuremath{\alpha}$ Clustering Shape $^{6}\mathrm{He}$},
  author = {Romero-Redondo, Carolina and Quaglioni, Sofia and Navr\'atil, Petr and others},
  journal = {Phys. Rev. Lett.},
  volume = {117},
  issue = {22},
  pages = {222501},  
  year = {2016},  
  publisher = {American Physical Society},
  doi = {10.1103/PhysRevLett.117.222501},
  url = {https://link.aps.org/doi/10.1103/PhysRevLett.117.222501}
}

@article{elhatisari_ab_2017,
    title = {\textit{{Ab} initio} Calculations of the Isotopic Dependence of Nuclear Clustering},
    volume = {119},
    doi = {10.1103/PhysRevLett.119.222505},
    url={https://link.aps.org/doi/10.1103/PhysRevLett.119.222505},
    journal = {Phys. Rev. Lett.},
    author = {Elhatisari, Serdar},
    year = {2017},
}

@article{song_ab_2026,
    title = {\textit{{Ab} initio} calculations of the carbon and oxygen isotopes: {Energies}, correlations, and superfluid pairing},
    volume = {872},
    issn = {0370-2693},
    shorttitle = {\textit{{Ab} initio} calculations of the carbon and oxygen isotopes},
    url = {https://www.sciencedirect.com/science/article/pii/S0370269325008445},
    doi = {10.1016/j.physletb.2025.140086},    
    journal = {Phys. Lett. B},
    author = {Song, Young-Ho and Kim, Myungkuk and Kim, Youngman and others},    
    year = {2026},
    pages = {140086},
}

@article{Calci2016,
    title = {Can Ab Initio Theory Explain the Phenomenon of Parity Inversion in $^{11}${Be}?},
    volume = {117},
    url = {https://link.aps.org/doi/10.1103/PhysRevLett.117.242501},
    doi = {10.1103/PhysRevLett.117.242501},  
    journal = {Phys. Rev. Lett.},
    publisher = {American Physical Society},
    author = {Calci, Angelo and Navrátil, Petr and Roth, Robert and others},    
    year = {2016},
    pages = {242501}
}

@article{shen_ab_2025,
    title = {Ab {Initio} {Study} of the {Beryllium} {Isotopes} $^{7}${Be} to $^{12}${Be}},
    volume = {134},
    url = {https://link.aps.org/doi/10.1103/PhysRevLett.134.162503},
    doi = {10.1103/PhysRevLett.134.162503},
    journal = {Phys. Rev. Lett.},
    publisher = {American Physical Society},
    author = {Shen, Shihang and Elhatisari, Serdar and Lee, Dean and others},    
    year = {2025},
    pages = {162503},
}

@article{Capel2021,
    title = {Combining {Halo}-{EFT} {Descriptions} of {Nuclei} and {Precise} {Models} of {Nuclear} {Reactions}},
    volume = {63},
    issn = {1432-5411},
    url = {https://doi.org/10.1007/s00601-021-01718-w},
    doi = {10.1007/s00601-021-01718-w},
    language = {en},
    number = {1},
    urldate = {2026-05-27},
    journal = {Few-Body Systems},
    author = {Capel, Pierre},
    month = dec,
    year = {2021},
    pages = {14},
}

\end{document}